\documentclass[aps,pra,twocolumn,superscriptaddress,showpacs]{revtex4-1}
\usepackage[version=3]{mhchem}
\usepackage[english]{babel}
\usepackage{dcolumn}
\usepackage{graphicx}
\usepackage{amsmath}
\usepackage{longtable}
\usepackage{bigstrut}
\usepackage{chemfig}
\usepackage{tabularx}
\usepackage{multirow}
\allowdisplaybreaks[1]
\def\bra#1{\mathinner{\langle{#1}|}} 
\def\ket#1{\mathinner{|{#1}\rangle}} 

\begin{document}

\title{Sensitivity of Transitions in Internal Rotor Molecules to a Possible Variation of the Proton-to-Electron Mass Ratio}

\author{Paul Jansen}
\affiliation{Institute for Lasers, Life and Biophotonics, VU University Amsterdam, 
De Boelelaan 1081, 1081 HV Amsterdam, The Netherlands}
\author{Isabelle Kleiner}
\affiliation{Laboratoire Interuniversitaire des Syst\`{e}mes Atmosph\'{e}riques (LISA), CNRS UMR 7583 et Universit\'{e}s Paris 7 et Paris Est, 61 avenue du G\'{e}n\'{e}ral de Gaulle, 94010 Cr\'{e}teil C\'{e}dex, France}
\author{Li-Hong Xu}
\affiliation{Department of Physics and Centre for Laser, Atomic, and Molecular Sciences, University of New Brunswick, Saint John, New Brunswick E2L 4L5, Canada}
\author{Wim Ubachs}
\affiliation{Institute for Lasers, Life and Biophotonics, VU University Amsterdam, 
De Boelelaan 1081, 1081 HV Amsterdam, The Netherlands}
\author{Hendrick L. Bethlem}
\affiliation{Institute for Lasers, Life and Biophotonics, VU University Amsterdam, 
De Boelelaan 1081, 1081 HV Amsterdam, The Netherlands}

\date{\today}

\begin{abstract}
Recently, methanol was identified as a sensitive target system to probe variations of the proton-to-electron mass ratio $\mu$ [Jansen \emph{et al.} Phys. Rev. Lett. \textbf{106}, 100801 (2011)]. The high sensitivity of methanol originates from the interplay between overall rotation and hindered internal rotation of the molecule -- i.e. transitions that convert internal rotation energy into overall rotation energy, or vice versa, give rise to an enhancement of the sensitivity coefficient, $K_{\mu}$. As internal rotation is a common phenomenon in polyatomic molecules, it is likely that other molecules display similar or even larger effects. In this paper we generalize the concepts that form the foundation of the high sensitivity in methanol and use this to construct an approximate model which allows to estimate the sensitivities of transitions in internal rotor molecules with $C_{3v}$ symmetry, without performing a full calculation of energy levels. We find that a reliable 
estimate of transition sensitivities can be obtained from the three rotational constants ($A$, $B$, and $C$) and three torsional constants ($F$, $V_3$ and $\rho$). This model is verified by comparing obtained sensitivities for methanol, acetaldehyde, acetamide, methyl formate and acetic acid with a full analysis of the molecular Hamiltonian. From the molecules considered, methanol appears to be the most suitable candidate for laboratory and cosmological tests searching for a possible variation of $\mu$. 
\end{abstract}


\maketitle

\section{Introduction\label{sec:introduction}}
Physical theories extending the Standard Model have presented scenarios that allow for, or even predict, spatio-temporal variations of the constants of nature~\cite{Uzan2003}. Currently, a number of laboratory experiments and astronomical observations are conducted to search for signatures of such variations. One of the dimensionless constants that are hypothesized to vary is the proton-to-electron mass ratio, $\mu=m_p/m_e$. A variation of $\mu$ can be detected by comparing frequencies of spectral lines in molecules as a function of time and/or position. A fractional change in $\mu$ will manifest itself as a fractional frequency shift. As a measure for the inherent sensitivity of a transition, the sensitivity coefficient, $K_{\mu}$, is defined by
\begin{equation}
\frac{\Delta \nu}{\nu}=K_{\mu}\frac{\Delta \mu}{\mu}.
\label{eq:Kmu}
\end{equation} 

For pure rotational transitions $K_{\mu}=-1$, for pure vibrational transitions $K_{\mu}=-\frac{1}{2}$, 
while for pure electronic transitions $K_{\mu}=0$~\cite{Varshalovich1993}. Transitions between the
inversion levels of ammonia \cite{Veldhoven2004,FlambaumNH3_2007} and hydronium (\ce{H3O+})\cite{Kozlov2011} have a sensitivity of $K_{\mu}=-4.2$ and $K_{\mu}=-2.5$, respecively. 
It was shown that the sensitivity of a transition between two near-degenerate levels that have a 
different functional dependence on $\mu$ is enhanced significantly~\cite{Flambaum2006,Flambaum&Kozlov2007,DeMille2008,Bethlem&Ubachs2008,Kozlov2011}. 
Recently, we reported such an enhancement for torsional-rotational transitions in 
methanol~\cite{Jansen2011}.

\begin{figure}[tb]
\includegraphics[width=\columnwidth]{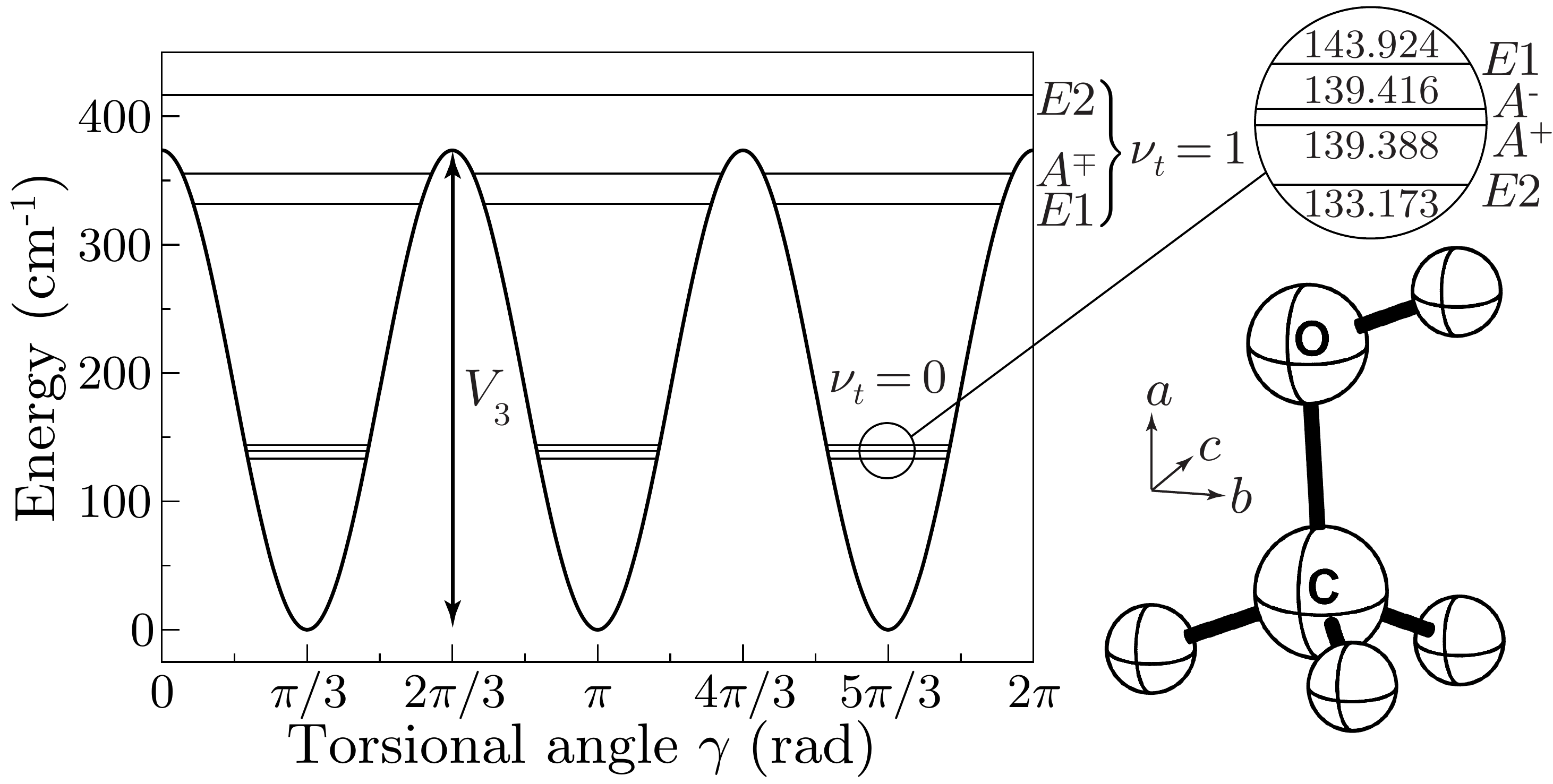}
\caption{Variation of the potential energy of methanol as function of the relative rotation $\gamma$ 
of the \ce{OH}-group with respect to the methyl group about the molecular axis. Shown are the $J = 1, |K| = 1$ 
energies of the lowest torsion-vibrational levels. The splitting between the different symmetry levels 
is due to tunneling through the potential barriers. The $A$-symmetry species are split further due to 
the asymmetry of the molecule ($K$-splitting).  
\label{fig:potential_and_structure}}
\end{figure}

Methanol (\ce{CH3OH}), schematically depicted in Fig.~\ref{fig:potential_and_structure}, 
consists of an OH group attached to a methyl group. The \ce{OH} and methyl group may rotate with respect to each other about the \ce{C-O} bond. On the left-hand side of the figure, the potential energy curve is shown as a function of the torsional angle, $\gamma$. The interaction between the \ce{OH} and methyl group results in a threefold barrier. Tunneling between the three wells, results in a splitting of each rotational level into three levels of different torsional symmetries~\cite{Lin&Swalen1959}. Transitions between the different torsional levels have a sensitivity coefficient, $K_{\mu}=-2.5$. As the torsional levels $A$ and $E$ belong to different symmetries, transitions between them are not allowed. It was shown in Jansen \emph{et al.}~\cite{Jansen2011} that transitions converting internal rotation energy into overall rotation energy, or vice versa, give rise to sensitivity coefficients, $K_{\mu}$, that range from -88 to +330 in the different isotopologues of methanol. 

Hindered internal rotation is a common phenomenon found in many polyatomic molecules. 
Hence, other molecules may have similar or larger sensitivities to a variation of $\mu$. In this paper 
we will calculate the sensitivities for methanol, acetaldehyde, acetamide, methyl formate and acetic 
acid, five relatively small molecules that have a group of $C_{3v}$ symmetry that rotates with respect to the remainder of the molecule. These five molecules have been detected in the interstellar medium of the local galaxy~\cite{Herbst2009} and some at high redshift~\cite{muller2011}. Methylamine, another relatively small internal rotor molecule is computationally more complex and will be treated in a separate paper~\cite{Ilyushin2011}.

This paper is organized as follows. In Section~\ref{sec:internalrotation}, we will give a brief review of the theory of internal rotor molecules and outline how the torsional-rotational levels are numerically calculated using the {\sc belgi} code~\cite{Hougen1994}. In addition, we present approximate expressions for obtaining the torsional energy splitting as function of the barrier height and the reduced moment of inertia, and compare this to the output of numerical calculations of a full Hamiltonian. In Section~\ref{sec:sensitivities}, we will discuss how the molecular constants that appear in the torsional-rotational Hamiltonian scale with $\mu$. These scaling relations are then used to determine the sensitivities of selected transitions in five different internal rotor molecules using {\sc belgi}. 
In Section~\ref{sec:toymodel} the analytical expressions for the torsional energy splitting presented 
in Section~\ref{sec:internalrotation} are used to construct a simple model for obtaining $K_{\mu}$ 
from the three rotational constants ($A$, $B$ and $C$) and three torsional constants ($F$, $V_3$ and $\rho$). This model provides an intuitive picture of the physics involved and makes it straightforward to estimate the sensitivity of other internal rotor molecules. 

\section{Hindered Internal Rotation\label{sec:internalrotation}}

A review of hindered internal rotation can be found in the seminal paper by Lin and 
Swalen~\cite{Lin&Swalen1959}, while a recent review of various effective Hamiltonians, methods and codes dealing with asymmetric-top molecules containing one internal rotor with $C_{3v}$ (or close to $C_{3v}$) symmetry can be found in the paper by Kleiner~\cite{Kleiner2010}. In this section we will summarize those results that are relevant for obtaining the sensitivity coefficients.  

\subsection{Hamiltonian}

The potential energy of an internal rotor molecule is a periodic function of the torsional angle $\gamma$ between the $C_{3v}$ group and the remainder of the molecule, as shown in Fig.~\ref{fig:potential_and_structure}. Hence, it can be expanded in a Fourier series as

\begin{equation}
V(\gamma)=\frac{V_3}{2}\left ( 1 - \cos{3\gamma} \right ) + \frac{V_6}{2}\left ( 1 - \cos{6\gamma} \right ) +
\ldots~~,
\label{eq:PotentialExpansion}
\end{equation}
where typically $V_6$ is about one hundred times smaller than $V_3$, but provides information on the shape of the torsional potential. If only the first term of the expansion is taken into account, the torsional wave functions and energies follow from the solutions of the Mathieu equation~\cite{Lin&Swalen1959}. 

\begingroup
\squeezetable
\centering
\renewcommand{\arraystretch}{1.25}
\begin{table*}[bt!]
\caption{Some low-order symmetry-allowed torsion-rotation ``Rho-Axis Method" (RAM) Hamiltonian terms for an asymmetric top containing a $C_{3v}$ 
internal rotor and a partial parameter list as used in the {\sc belgi} code (taken from Kleiner~\cite{Kleiner2010}). 
The $\mu$-dependence of the molecular constants is given in parenthesis.  
\label{tab:highorderterms}}
\begin{tabularx}{\textwidth}{@{}l l |>{\centering}X >{\centering}X >{\centering}X >{\centering}X >{\centering}X >
{\centering}X >{\centering\arraybackslash}X@{}}
\hline\hline
\bigstrut & & \multicolumn{7}{l}{Torsional/Potential}\\\cline{3-9}
\bigstrut & & 1  & $1-\cos{3\gamma}$ & $p_{\gamma}^2$  & $P_ap_{\gamma}$ & $1-\cos{6\gamma}$ & $p_{\gamma}^4$ & $P_ap_{\gamma}^3$ \\
Rotational & & $(\mu^{0})$ & $(\mu^{0})$ & $(\mu^{-1})$ & $(\mu^{0})$ & $(\mu^{0})$ & $(\mu^{-2})$ & $(\mu^{-1})$\\[0.25ex]\hline
1 & $(\mu^0)$ & & $V_3/2$  & $F$ & $\rho$ & $V_6/2$ & $k_4$ & $k_3$ \\
$P^2$ & $(\mu^{-1})$ & $(B+C)/2$  & $F_v$& $G_v$& $L_v$& $N_v$& $M_v$& $k_{3J}$\\
$P_a^2$ & $(\mu^{-1})$ & $A-(B+C)/2$ & $k_5$ & $k_2$ & $k_1$ & $K_2$& $K_1$& $k_{3K}$\\
$P_b^2-P_c^2$ & $(\mu^{-1})$& $(B-C)/2$& $c_2$& $c_1$& $c_4$& $c_{11}$& $c_3$& $c_{12}$\\
$P_aP_b+P_bP_a$ & $(\mu^{-1})$& $D_{ab}$& $d_{ab}$& $\Delta_{ab}$& $\delta_{ab}$ & $d_{ab6}$ & ${\Delta\Delta}_{ab}$ & ${\delta\delta}_{ab}$\\[0.25ex]
\hline\hline
\end{tabularx}
\end{table*}
\endgroup

To express the kinetic part of the Hamiltonian an axis system $(a,b,c)$ must be defined. The symmetric part of the molecule is defined as the internal rotor or top whereas the remainder of the molecule is referred to as the frame, which for all molecules presented here has a plane of symmetry. The origin of the coordinate system coincides with the center of mass of the molecule. The $a$ axis is chosen parallel to the symmetry axis of the top, and the $b$ axis lies in the plane of symmetry. The $c$ axis follows from the definition of a right-handed coordinate system. The inertia tensor then takes the form

\begin{equation}
\mathbf{I}=
\begin{bmatrix}
I_{c} 	& 0		& 0	 	\\
0			& I_{b}	& -I_{ab}	\\
0			& -I_{ab}	& I_{a}  \\
\end{bmatrix}
,
\label{eq:InertiaTensor}
\end{equation}
\noindent
with $I_a=\sum_i{m_i(b_i^2+c_i^2)}$ being the moment of inertia of the molecule about the $a$ axis. The subscript $i$ runs over all atoms with mass $m_i$ in the molecule. Expressions for $I_b$ and $I_c$, the moments of inertia around the $b$ and $c$ axis, respectively, can be found by cyclic permutation of the $a$, $b$ and $c$ labels. $I_{ab}=\sum_i{m_ia_ib_i}$ is the product of inertia about the $a$ and $b$ axis. The kinetic energy can be expressed 
as~\cite{Lin&Swalen1959}
\begin{multline}
T=\frac{1}{2}I_a\omega_a^2 + \frac{1}{2}I_b\omega_b^2 + \frac{1}{2}I_c\omega_c^2\\ - I_{ab}\omega_a\omega_b + \frac{1}{2}I_{a2}\dot{\gamma}^2 + I_{a2}\omega_a \dot{\gamma},
\label{eq:KineticEnergy}
\end{multline}
\noindent
with $\omega_a$, $\omega_b$, and $\omega_c$ the angular velocity components around the $a$, $b$, and $c$ axis, respectively, and $I_{a2}$ the moment of inertia of the top along its own symmetry axis. For a given vibrational state the zeroth order torsion-rotation Hamiltonian can be separated into a symmetric top part, an asymmetric top part and a torsional part~\cite{Lees&Baker1968,Herbst1984}

\begin{subequations}
\label{eq:fullH}
\begin{align}
H^0 &= H_{\text{RS}}^0+H_{\text{RA}}^0+H_{\text{tors}}^0,\\
\intertext{where}
H_{\text{RS}}^0  &= \frac{1}{2}\left ( B+C \right ) \left ( P_b^2 + P_c^2 \right ) + AP_a^2 \label{eq:Hsym}\\
H_{\text{RA}}^0 &=  \frac{1}{2}\left ( B-C \right ) \left ( P_b^2 - P_c^2 \right ) + D_{ab}\left (P_aP_b+P_bP_a \right )\label{eq:Hasym}\\
H_{\text{tors}}^0  &=  F \left ( p_{\gamma} + \rho P_a\right )^2 + V(\gamma)\label{eq:Htors}.
\end{align}
\end{subequations}
\noindent
$P_a,P_b$ and $P_c$ are the usual angular momentum operators along the $a, b$ and $c$ axis, respectively, and $p_{\gamma}=-i\partial/\partial\gamma$ is the angular momentum operator associated with the internal rotation of the top with respect to the frame. The coupling between the internal rotation and overall rotation in Eq.~\eqref{eq:Htors} can be eliminated partly by transforming to a different axis system, the so-called ``Rho-Axis System". In the resultant ``Rho-Axis Method" (RAM), which is implemented in the {\sc belgi} code used hereafter, the torsional Hamiltonian operator contains only the $+2F\rho P_a {p_{\gamma}}$ term. It is important to note here that two sign conventions exist in the literature for the torsion-rotation operator in Eq.~\eqref{eq:Htors}, i.e., $F(p_{\gamma}+\rho P_a)^2$ and $F(p_{\gamma}-\rho P_a)^2$. If the latter convention is adopted the $\pm K$ labeling of the $E$ levels (\emph{vide infra}) is reversed~\cite{Ilyushin2001}. In this paper we adopt the convention with the ``+" sign - i.e. $F(p_{\gamma}+\rho P_a)^2$. The effective rotational and torsional constants are defined by

\begin{align}
A &= \frac{1}{2} \hbar^2 \left ( \frac{I_a+I_b}{I_aI_b-I_{ab}^2} - \frac{I_b}{I_b^2+I_{ab}^2} \right )\label{eq:A}, \\
B &= \frac{1}{2} \hbar^2 \frac{I_b}{I_b^2+I_{ab}^2}\label{eq:B},\\
C &= \frac{1}{2}\hbar^2 \frac{1}{I_c} \label{eq:C},\\
D_{ab} &= \frac{1}{2} \hbar^2 \frac{I_{ab}}{I_b^2+I_{ab}^2}\label{eq:Dab},\\
F &= \frac{1}{2} \hbar^2 \frac{I_aI_b-I_{ab}^2}{I_{a2}\left ( I_{a1}I_b - I_{ab}^2 \right )}\label{eq:F},
\end{align}

\noindent
where $I_{a1}$ is the moment of inertia attributed to the frame, defined by $I_{a1}=I_a-I_{a2}$. 
A dimensionless parameter $\rho$ is introduced by the axis transformation described earlier. For a symmetric top, $\rho$ is simply defined as the ratio between the moment of inertia of the top divided by the moment of inertia of the molecule along the $a$-axis, i.e. $\rho=I_{a2}/I_a$. For asymmetric molecules $\rho$ is a more complicated function of the various moments of inertia:
\begin{equation}
\rho = \frac{I_{a2}\sqrt{I_b^2+I_{ab}^2}}{I_aI_b-I_{ab}^2}\label{eq:rho}.
\end{equation}

The Hamiltonian of Eq.~\eqref{eq:fullH} can be expanded by adding additional distortion and interaction terms. Many of these higher-order constants multiply torsional operators by rotational operators and can be considered as effective constants after the van Vleck transformations of the torsion-rotation 
Hamiltonian~\cite{kirtman1962,Herbst1984,Kleiner2010}. Some low-order symmetry-allowed torsion-rotation terms for an asymmetric top containing a $C_{3v}$ internal rotor are listed in Table~\ref{tab:highorderterms}.

The overall Hamiltonian can now be written as
\begin{equation}
H = H^0 + H_{\text{int}} + H_{\text{c.d.}}\label{eq:overallH},
\end{equation}
\noindent
where $H_{\text{c.d.}}$ corresponds to the centrifugal distortion Hamiltonian, and $H_{\text{int}}$ contains higher-order torsional-rotation interaction terms. 

\subsection{Eigenfunctions and Eigenvalues}
\subsubsection{Torsion}
Herbst \emph{et al.}~\cite{Herbst1984} suggested to evaluate the Hamiltonian of Eq.~\eqref{eq:overallH} in two steps. In the first step, the torsional Hamiltonian $H_{\text{tors}}^0$ is diagonalized in a product basis set composed of free rotor torsional eigenfunctions of $p_{\gamma}$ and eigenfunctions $\ket{K}=(2\pi)^{-1/2}\exp(-i{K}\chi)$ of $P_a$, where $\chi$ is the Euler angle: 
\begin{equation}
\ket{Kk\sigma} = \frac{1}{\sqrt{2\pi}} \ket{K} \exp{\left ( i \left [ 3k+\sigma \right ] \gamma \right )},
\label{eq:torsbasis}
\end{equation}
\noindent
where $\sigma$ can take the values $-1$, $0$, or $+1$ and $k$ can be any integer. The eigenvalues of $p_{\gamma}$ are $3k+\sigma$ as required by the periodicity of the potential. Due to the symmetry of the torsional Hamiltonian, basis functions of different $\sigma$ do not mix. Moreover, basis functions of different $K$ do not mix either, as $P_a$ and $H_{\text{tors}}^0$ commute. The resulting Hamiltonian matrix for each value of $\sigma$ and $K$ is infinite in size, but it was found that truncating it to a $21\times 21$ matrix ($-10\leq k \leq 10$) is sufficient to obtain experimental accuracy for the molecules under study here~\cite{Herbst1984}. The torsional eigenfunctions can be written as

\begin{equation}
\ket{K\nu_t\sigma}=\frac{1}{\sqrt{2\pi}}\ket{K}\sum_{k=-10}^{10}{A_{3k+\sigma}^{K,\nu_t}\exp{\left ( i \left [ 3k+\sigma \right ] \gamma \right )}},
\label{eq:basisset}
\end{equation}

\noindent
where $\nu_t$ is the torsional vibration quantum number and $A_{3k+\sigma}^{K,\nu_t}$ are expansion coefficients. States with $\sigma=0$ are labeled as $A$, and states with $\sigma=+1$ and $\sigma=-1$ are labeled as $E1$ and $E2$, respectively. For $A$ torsional states, $\pm{K}$ levels are degenerate, whereas for $E$ states a degeneracy exists between $E1$, $K$ and $E2$, $-K$ levels. Although the torsional $E1$ and $E2$ state have different labels, transitions between these two states are allowed. It was therefore suggested by Lees~\cite{Lees1973} to refer to $E1$ and $E2$ levels as $E$ levels where the sign of $K$ distinguishes the two symmetries.   

\begin{figure}[bt]
\includegraphics[width=\columnwidth]{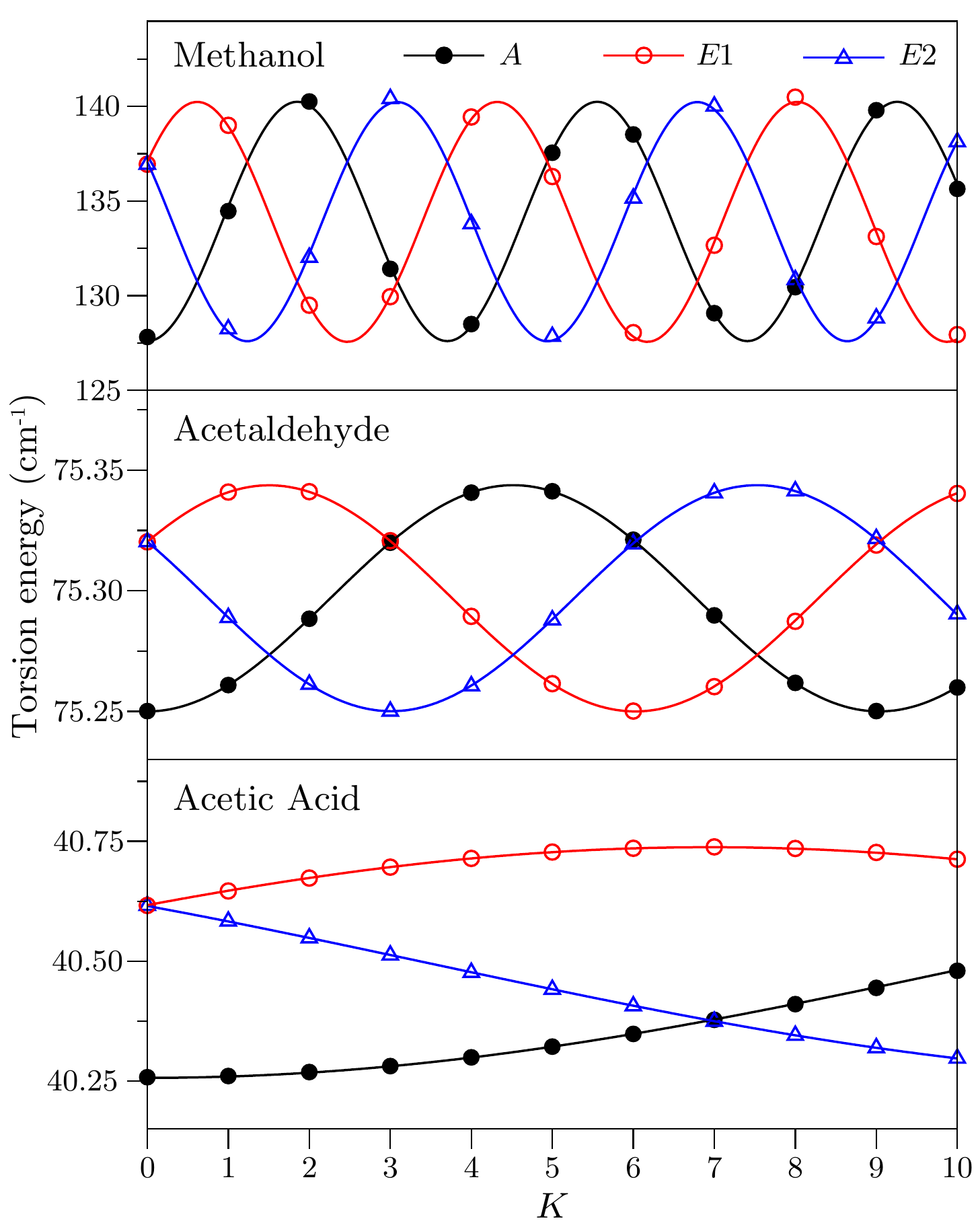}
\caption{Torsional energies obtained with {\sc belgi} for $A$ (solid circles), $E1$ (open circles), and $E2$ (open triangles) levels as function of $K$ for methanol \ce{(CH3OH)}, acetaldehyde \ce{(CH3COH)}, and acetic acid \ce{(CH3COOH)} for $\nu_t=0$. The solid curves are fits to Eq.~\eqref{eq:Etors} for $A$, $E1$, and $E2$ states. Note that only integer values of $K$ have physical meaning. 
\label{fig:torsional_energies}}
\end{figure}

The eigenvalues of the torsional Hamiltonian for methanol, acetaldehyde and acetic acid are depicted in 
Fig.~\ref{fig:torsional_energies} as a function of $K$ for the ground torsional state ($\nu_t=0$) of those molecules. The solid circles, open circles and open triangles represent values numerically calculated using the {\sc belgi} code~\cite{Hougen1994,Xu2008o,Kleiner1996,Ilyushin2008} for the $A$, $E1$ and $E2$ torsional states, respectively. It is seen that the torsional energies are periodic functions of $K$ with a period that is proportional to $\rho^{-1}$ ($\rho=0.81$, $0.33$ and $0.07$ for methanol, acetaldehyde and acetic acid, respectively).

\begin{figure}[tb]
\includegraphics[width=\columnwidth]{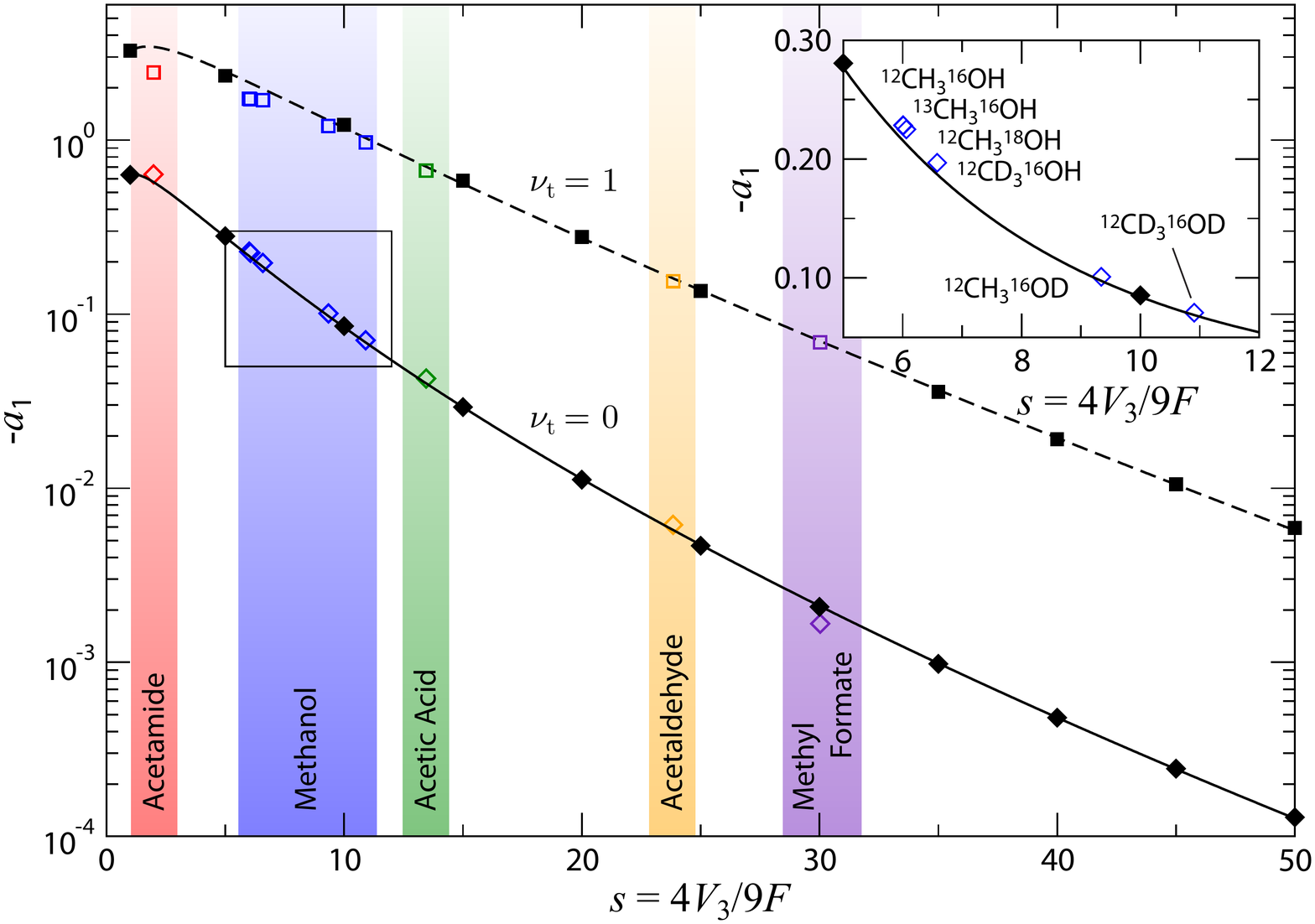}
\caption{The $a_1$ expansion coefficients of Eq.~\eqref{eq:Etors} as a function of the reduced barrier height $s$ for the ground torsional state $\nu_t=0$ and first excited torsional state $\nu_t=1$. Solid diamonds and squares represent $a_1$ coefficients in $\nu_t=0$ and $\nu_t=1$, respectively, determined by fitting the eigenvalues of Eq.~\eqref{eq:Htors} obtained with {\sc belgi} for different values of $s$ according to the expansion of Eq.~\eqref{eq:Etors}. The solid and dashed curves are fits according to Eq.~\eqref{eq:a1} and Eq.~\eqref{eq:a1vt=1}, respectively. Open diamonds and squares are $a_1$ expansion coefficients for several molecules taking into account higher-order torsional parameters. The inset shows an enlargement of the $\nu_t=0$ curve near the values for the six isotopologues of methanol. 
\label{fig:expansion_coefficients}}
\end{figure}

In order to obtain an analytical model for estimating the sensitivities of transitions in internal rotor molecules, to be discussed in Section~\ref{sec:toymodel}, we will now derive approximate solutions to the torsional Hamiltonian. It is clear from Eq.~\eqref{eq:basisset} that substituting $\sigma$ with $\sigma+3$ results in identical eigenvalues, consequently, the eigenvalues may be regarded as periodic functions which can be expanded in a Fourier series as~\cite{Lin&Swalen1959}

\begin{equation}
E_{\text{tors}}=F \left [ a_0 + a_1 \cos{\left \{\frac{2\pi}{3} \left ( \rho{K}+\sigma \right )\right \}}+\ldots \right ],
\label{eq:Etors}
\end{equation}
\noindent
where $a_0, a_1$ and higher order terms are (dimensionless) expansion coefficients. It can be shown~\cite{Lin&Swalen1959} that these coefficients are functions of the reduced barrier height $s$, with
\begin{equation}
s=\frac{4V_3}{9F},
\label{eq:reducedbarrier}
\end{equation}
\noindent
and that in the moderate to high-barrier limit, the series converges quickly. The solid curves shown in Fig.~\ref{fig:torsional_energies} are obtained by fitting the first two terms of Eq.~\eqref{eq:Etors} for $A$ $(\sigma = 0)$, $E1$ $(\sigma = +1)$, and $E2$ $(\sigma = -1)$ states. The resulting coefficients for methanol, acetaldehyde and acetic acid, as well as those for acetamide and methyl formate are plotted as the open diamonds in Fig.~\ref{fig:expansion_coefficients}. By diagonalizing $H_{\text{tors}}^0$ for several values of $s$ and $\rho>0$, while all other constants are set to zero, and fitting the torsional energies according to Eq.~\eqref{eq:Etors}, $a_1$ coefficients were obtained for each value of $s$. These generic coefficients are plotted as the solid diamonds in Fig.~\ref{fig:expansion_coefficients}. According to Lin and Swalen~\cite{Lin&Swalen1959} the $a_1$ coefficients are given by the following equation:

\begin{equation}
a_1 = A_1{s^{B_1}}\cdot e^{-C_1\sqrt{s}}.\label{eq:a1}
\end{equation} 
\noindent
The solid line shown in Fig.~\ref{fig:expansion_coefficients} is obtained by fitting Eq.~\eqref{eq:a1} to the generic $a_1$ values, using $A_1=-5.296, B_1=1.111$ and $C_1=2.120$. Note that these fit parameters deviate from those given in Table~IV of Ref.~\cite{Lin&Swalen1959}, but the curves agree with the curves shown in Fig~7 of the same paper. Small differences between the curves and the $a_1$ coefficients obtained for the different molecules can be attributed to higher order torsional terms which were not taken into account to obtain the fits. 

\begin{figure*}[tb]
\includegraphics[width=1\textwidth]{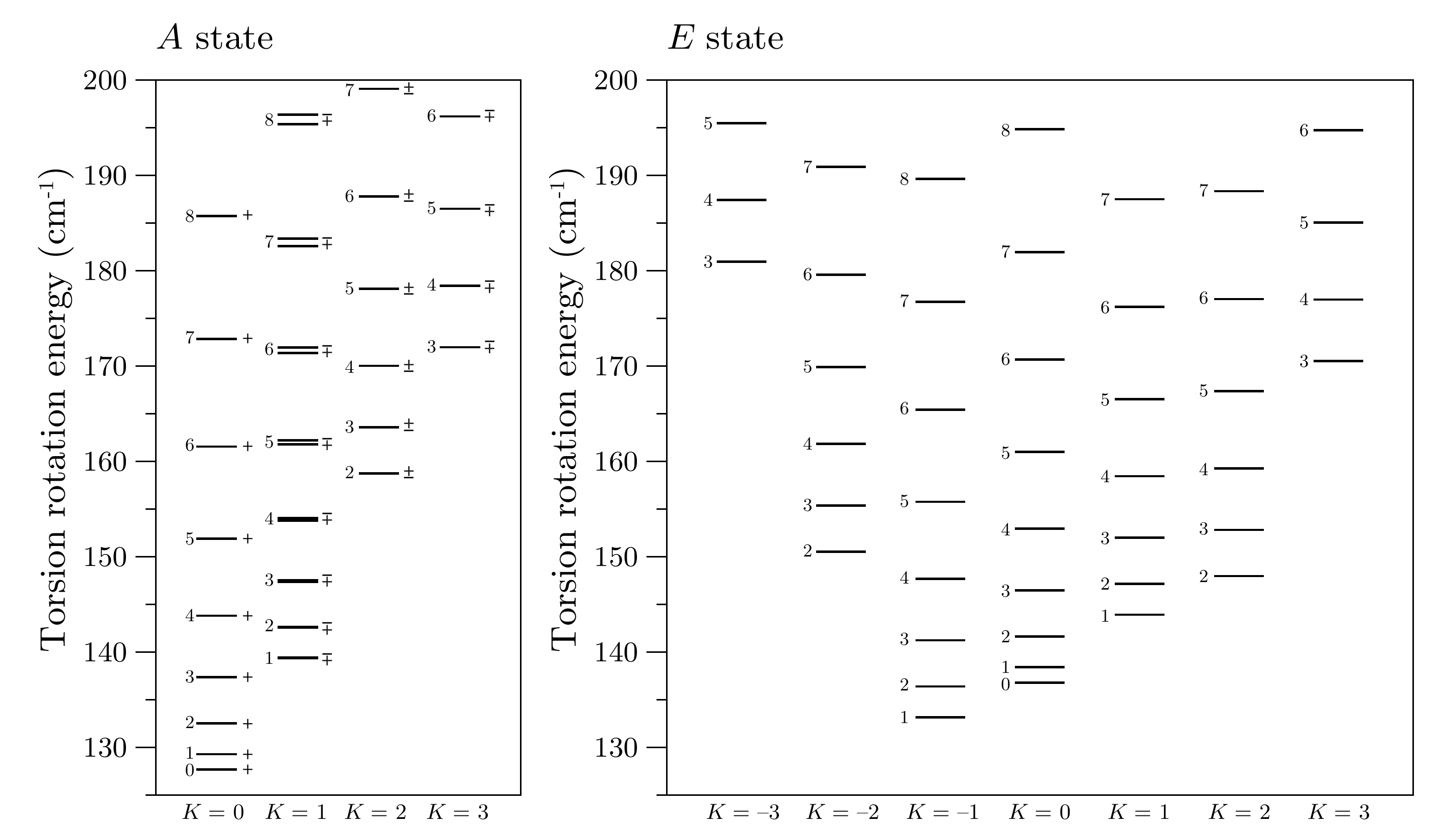}
\caption{Energy of the lowest rotational levels in the torsion-vibrational groundstate ($\nu_{t} = 0$) of methanol \ce{(^{12}CH3^{16}OH)}~\cite{Xu2008o}. The levels are denoted by $J$ (indicated on the left side of each level), $K$. For the $A$ levels the so-called parity quantum number ($+/-$) is also used. The panel on the left contains the levels for the $A$ state, whereas the panel on the right contains the levels for the $E$ state. High sensitivities are expected for transitions that connect near degenerate levels with different $K$. 
\label{fig:K-ladders}}
\end{figure*}

\subsubsection{Rotation}
At this point the torsional Hamiltonian is diagonalized and the first step of the approach by Herbst \emph{et al.} is complete. The second step of the approach consists of evaluating the remainder of the full Hamiltonian, i.e., overall rotation and coupling terms, in the basis set~\cite{Herbst1984}

\begin{equation}
\ket{JK\nu_t\sigma}=\ket{JK}\ket{K\nu_t\sigma}\label{eq:totalbasis},
\end{equation}
\noindent
with $\ket{JK}$ the symmetric top rotational eigenfunctions. The eigenvalues of the Hamiltonian of Eq.~\eqref{eq:Hsym} are
\begin{multline}
\bra{JK\nu_t\sigma}H_{\text{Rsym}}^0\ket{JK\nu_t\sigma}=\\
\frac{1}{2}\left ( B+C \right )J\left ( J+1 \right )+\left ( A-\frac{B+C}{2} \right )K^2\label{eq:rotEn}.
\end{multline}

\noindent
Note that, in the case of an asymmetric top, torsional $A$ levels are split in $\pm$ components for $K>0$ due to the asymmetry of the molecule. 

\subsection{Level Schemes and Selection Rules}

In Fig.~\ref{fig:K-ladders} the lowest energy levels of \ce{^{12}CH3^{16}OH} are shown for the $A$ and $E$ species. $A$ and $E$ symmetry species can be considered  as two different molecular species in the same sense as para- and ortho-hydrogen: radiative transitions beween $A$ and $E$ species do not occur. 
The arrangement of energy levels within a symmetry state is quite similar to the structure of the $K$ ladders in a (prolate) symmetric top, however in the case of internal rotation each $K$ ladder attains an additional offset, $E_{\text{tors}}(K)$, induced by the tunneling splitting. The + or - component of the $A$ state refers to the $\ket{J,K,\nu_t,0}\pm\ket{J,-K,\nu_t,0}$ and $\ket{J,K,\nu_t,0}\mp\ket{J,-K,\nu_t,0}$ linear combinations of basis functions for $K$ even and $K$ odd respectively. The overall parity of the levels, is given by $\pm(-1)^{J+\nu_t}$~\cite{Herbst1984}. When transitions between levels of different torsion-vibration 
states are ignored, the selection rules for allowed transitions are~\cite{Herbst1984}

\begin{alignat*}{4}
A\text{ levels:}\quad &\pm \leftrightarrow \mp \quad  & &\Delta{J} &=0  \quad  & |\Delta{K}|&=&\; 0\, (K\neq 0),1\\
                      &\pm \leftrightarrow \pm \quad  &|&\Delta{J}|&=1  \quad  & |\Delta{K}|&=&\; 0,1\\
E\text{ levels:}\quad &                               & &\Delta{J} &=0  \quad  & |\Delta{K}|&=&\; 1\\
                      &                               &|&\Delta{J}|&=1  \quad  & |\Delta{K}|&=&\; 0,1
\end{alignat*}
where $K$ is only a good quantum number in the limit of a symmetric top. As a consequence, transitions with $|\Delta{K}|>1$ are allowed in asymmetric top molecules.    

\section{Scaling and Sensitivity Coefficients\label{sec:sensitivities}}

The sensitivity to a variation of $\mu$ of a transition between states 
$\vert \nu_{t}'',J'',K'',\textit{Ts}'' \rangle$ and $\vert \nu_{t}',J',K',\textit{Ts}'\rangle$ is given by
\begin{multline}
K_{\mu} \left( \nu_{t}'',J'',K'',\textit{Ts}'' \rightarrow \nu_{t}',J',K',\textit{Ts}' \right) = \\
\frac{\mu \left( {\partial E}\middle/{{\partial \mu}} \right)_{\nu_{t}'',J'',K'',\textit{Ts}''} 
- \mu \left( {\partial E}\middle/{\partial \mu} \right)_{\nu_{t}',J',K',\textit{Ts}'}}
{E(\nu_{t}'',J'',K'',\textit{Ts}'')-E(\nu_{t}',J',K',\textit{Ts}')}.
\label{eq:K_resonance}
\end{multline}
Thus, in order to calculate the $K_{\mu}$ coefficients, the energy of each level and 
its dependence on $\mu$ has to be obtained. This translates into knowing the values of the molecular constants that go into {\sc belgi} and how these constants scale with $\mu$.

We will first examine the scaling relations for the lowest order constants. We implicitly assume that the neutron-electron mass ratio follows the same behavior as the proton-electron mass ratio, and that no effects depending on quark structure persist~\cite{Ubachs2007}. As a consequence, the atomic masses and hence the moments of inertia are directly proportional to $\mu$. The rotational, centrifugal and torsional constants $A$, $B$, $C$, $D_{ab}$, and $F$ and the factor $\rho$ are explicit functions of $I_a$, $I_b$, $I_c$, and $I_{ab}$, their $\mu$ dependence is obtained from Eq.~\eqref{eq:A}--\eqref{eq:rho}. Within the Born-Oppenheimer approximation the torsional potential $V_3$ is independent of the mass of the nuclei and hence of $\mu$. 

The scaling relations for higher order constants are derived from multiple combinations of lower-order torsional and rotational operators. Some symmetry-allowed torsion-rotation terms are listed in Table~\ref{tab:highorderterms}. Let us, for example, inspect the constant $M_{\nu}$, 
which can be considered as a product of the torsional operator $p_{\gamma}^4$ with the rotational operator $P^2$. As the term with $p_{\gamma}^4$ scales as $\mu^{-2}$ and the term with $P^2$ scales as $\mu^{-1}$, we expect $M_{\nu}$ to scale as $\mu^{-3}$. The scaling relations of constants associated with other operators follow in a similar manner. The supplementary material of Ref.~\cite{Jansen2011} lists the scaling relations for all constants used for methanol.  

\begin{figure}[tb]
\includegraphics[width=\columnwidth]{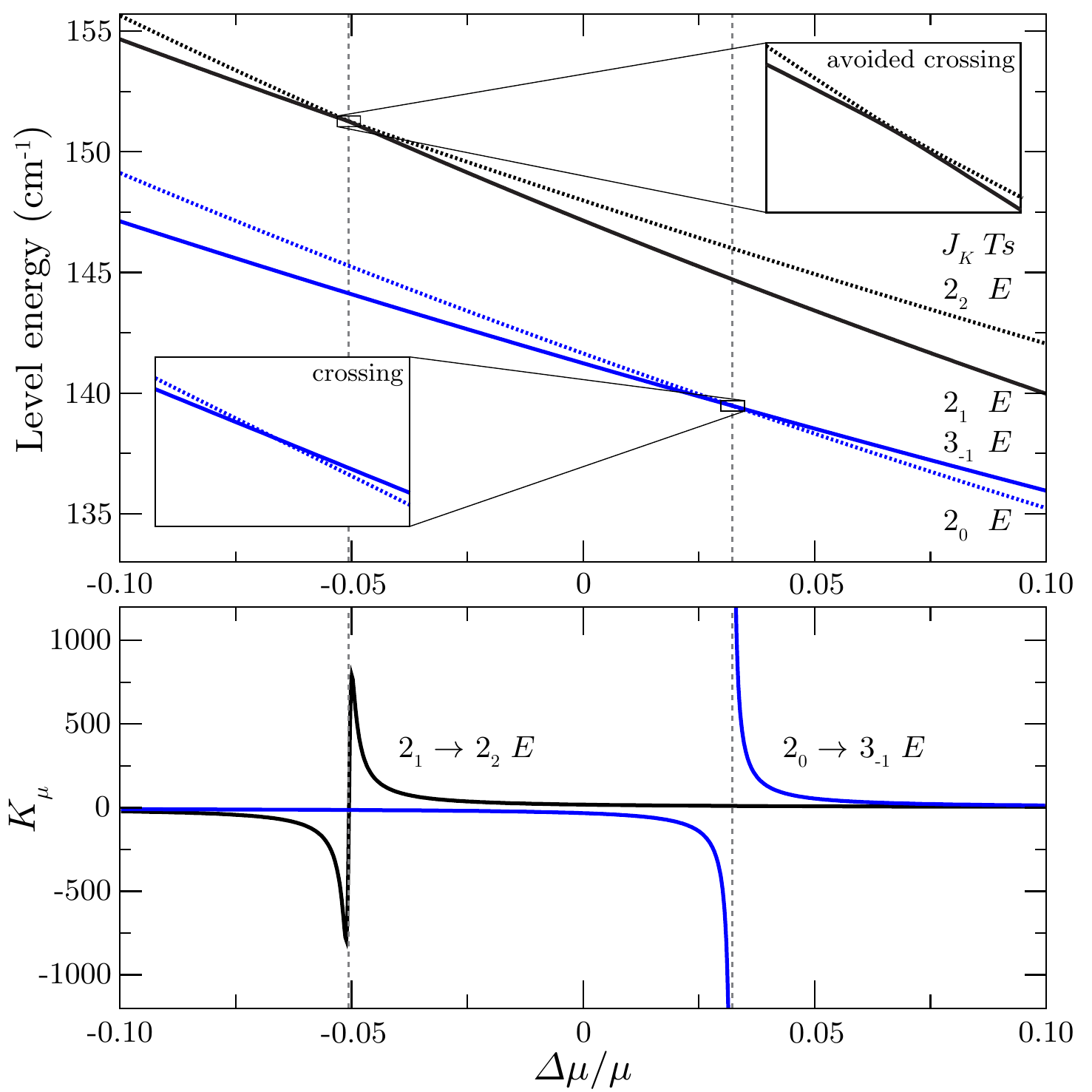}
\caption{Energies of selected rotational levels in methanol \ce{(^{12}CH3^{16}OH)} as a function of the 
fractional change in the proton-to-electron mass ratio. The insets in the upper panel show magnifications 
of the regions where the levels cross. It can be seen that levels with same $J$ give rise to avoided crossings, whereas levels with different $J$ do not. The lower panel depicts the sensitivity coefficients of the $2_1 \rightarrow 2_2\,E$ and $2_{0}\rightarrow 3_{-1}\,E$ transitions as a function of $\Delta\mu/\mu$.
For transitions between levels with same $J$, $K_{\mu}$ goes to zero at the (avoided) crossing, whereas 
transitions between levels with different $J$ $K_{\mu}$ diverges to infinity at the crossing. 
\label{fig:avoided_crossings}}
\end{figure}

In order to determine the sensitivity coefficients, we have written a computer code that 
generates the molecular constants as a function of $\mu$ using the discussed scaling relations, 
calls {\sc belgi} with these constants as input, and subsequently stores the computed level energies. 

As an example, the upper panel of Fig.~\ref{fig:avoided_crossings} shows the energies of the 
$2_2\,E$, $2_1\,E$, $3_{-1}\,E$ and $2_0\,E$ levels in methanol as a function of $\Delta\mu/\mu$. 
The sensitivity coefficient of the $2_0 \rightarrow 3_{-1}\,E$ and 
$2_1 \rightarrow 2_2\,E$ transitions can be obtained by dividing the difference 
in slope by the difference in energy, \emph{cf.} Eq.~\eqref{eq:K_resonance}. The sensitivity 
coefficients of these transitions as a function of the relative variation of the proton-electron mass ratio, $\Delta\mu/\mu$, are shown in the lower panel of Fig.~\ref{fig:avoided_crossings}. As expected from Eq.~\eqref{eq:K_resonance}, the sensitivity is strongly enhanced when the energy difference between the levels becomes small. 

The behavior of the sensitivity coefficient close to the resonance depends on the quantum numbers of the levels involved. For instance, in methanol, $K$ is not a good quantum number due to the asymmetry of the molecule, and levels with equal $J$ mix. As a consequence, the $2_2\,E$ and $2_1\,E$ levels shown in Fig.~\ref{fig:avoided_crossings} display an avoided crossing, and the sensitivity coefficient for the $2_1 \rightarrow 2_2\,E$ transition is zero at the resonance. In contrast, $J$ is a good quantum number and levels of different $J$ will not mix. As a consequence, the sensitivity coefficient for the $2_0 \rightarrow 3_{-1}\,E$ transition becomes infinite at the resonance. In practice, we are only interested in the value of $K_{\mu}$ at $\Delta\mu/\mu<10^{-5}$ and the effects of avoided crossings is relevant only if the levels cross extremely close to $\Delta\mu/\mu=0$. In our study we have only come across one transition that has a significantly reduced sensitivity coefficient as a result of mixing of the energy levels involved, namely the $8_{0} \rightarrow 8_{-1} E$ transition in \ce{^{12}CD3^{16}OD} at 4.2\,GHz with $K_{\mu}=0.7$. Note that the sign of the sensitivity coefficients at $\Delta\mu/\mu=0$ is positive if the levels cross at $\Delta\mu/\mu<0$, and negative if the levels cross at $\Delta\mu/\mu>0$.  

\begingroup
\squeezetable
\centering
\begin{table*}[htb!]
\caption{Selected transitions~\cite{Lovas2004} and $K_\mu$ coefficients for different molecules, calculated with {\sc belgi} (fourth column) using the constants from Refs.~\cite{Xu2008o,Kleiner1996,Ilyushin2004,Carvajal2007,Ilyushin2008} and the approximate model (fifth column) discussed in the text. For methanol the most sensitive lines are shown, whereas for the other molecules the eight lowest transitions that have been observed in the interstellar medium as listed in the review by Lovas~\cite{Lovas2004} are given. The error in the last digit(s) is quoted within brackets. Molecules marked with an asterisk (*) are labeled according to the sign convention opposite to the one as used in the text to be consistent with literature -- i.e. $F(p_{\gamma}-\rho P_a)^2$ instead of $F(p_{\gamma}+\rho P_a)^2$. As a consequence the sign of the $\pm K$ labeling of the $E$ transitions is negated for these molecules.}  
\label{tab:sensitivities}
\begin{tabular*}{1\textwidth}{@{\extracolsep{\fill}}l l D..{4} D..{5} D..{5}}
\toprule
Molecule &
{Transition, $J_K$ }&
\multicolumn{1}{c}{Transition (MHz)}&
\multicolumn{1}{c}{$K_{\mu}^{\text{{\sc belgi}}}$} &
\multicolumn{1}{c}{$K_{\mu}^{\text{toy}}$} \bigstrut \\
\hline
\\[-1.8ex]
Methanol
                     & $   5_{   1} \rightarrow    6_{   0} A^{ + }$ &         6\,668.5192(8)		   &     -42.(2)      & -46     \\
                     & $   9_{  -1} \rightarrow    8_{  -2} E      $ &         9\,936.202(4) 			&      11.5(6)     &  16.7   \\
                     & $   5_{   2} \rightarrow    4_{ 3} A^{+}    $ &         9\,978.686(4)          &        53.(3)    &  35     \\
                     & $   5_{   2} \rightarrow    4_{ 3} A^{-}    $ &        10\,058.257(12)   	   &       52.(3)     &  35     \\
                     & $   2_{   0} \rightarrow    3_{  -1} E      $ &        12\,178.593(4) 			&     -33.(2)      & -32     \\
                     & $   2_{   1} \rightarrow    3_{   0} E      $ &        19\,967.396(2) 			&      -5.9(3)     &  -5.0   \\
                     & $   9_{   2} \rightarrow   10_{   1} A^{ + }$ &        23\,121.024(2) 			&     -11.7(6)     & -10.8   \\
                     & $   3_{   2} \rightarrow    3_{   1} E      $ &        24\,928.715(14) 			&      17.9(9)     &  15.2   \\
                     & $   2_{   2} \rightarrow    2_{   1} E      $ &        24\,934.382(5) 			&      17.9(9)     &  15.2   \\
                     & $   8_{   2} \rightarrow    9_{   1} A^{ - }$ &        28\,969.954(20) 			&      -9.5(6)     &  -8.8   \\
                     & $   4_{  -1} \rightarrow    3_{   0} E      $ &        36\,169.290(14) 			&       9.7(5)     &   9.6   \\
                     & $   6_{   2} \rightarrow    5_{   3} A^{ - }$ &        38\,293.292(14) 			&     -15.1(8)     & -10.4   \\
                     & $   6_{   2} \rightarrow    5_{   3} A^{ + }$ &        38\,452.653(14) 			&     -15.0(8)     & -10.4   \\
                     & $   7_{   0} \rightarrow    6_{   1} A^{ + }$ &        44\,069.476(15) 			&       5.2(3)     &   5.9   \\
                     & $   1_{   0} \rightarrow    2_{  -1} E      $ &        60\,531.489(10) 			&      -7.4(4)     &  -7.3   \\
                     & $   1_{   1} \rightarrow    2_{   0} E      $ &        68\,305.680(7) 			&      -2.4(1)     &  -2.2   \\[1.5mm]
Acetaldehyde        & $   1_{-1}\rightarrow 1_{1} E               $ &         1\,849.634(7)          &      -3.7(2)     &  -4.2   \\
                     & $   1_{1}\rightarrow 2_{0 A^{+}}$             &         8\,243.462(3)          &      -1.11(6)    &  -1.15  \\
                     & $ 1_0 \rightarrow 0_0 E $                     &        19\,262.140(4)          &      -1.00(5)    &  -1.00  \\
                     & $ 1_0 \rightarrow 0_0 A^+$                    &        19\,265.137(1)          &      -1.00(5)    &  -1.00  \\
                     & $ 2_0 \rightarrow 1_0 E$                      &        38\,506.035(3)          &      -1.00(5)    &  -1.00  \\
                     & $ 2_0 \rightarrow 1_0 A^+$                    &        38\,512.081(3)          &      -1.00(5)    &  -1.00  \\
                     & $ 1_{-1} \rightarrow 1_0 E$                   &        47\,746.980(5)          &      -1.03(5)    &  -0.93  \\
                     & $ 1_1 \rightarrow 1_0 A^+$                    &        47\,820.620(4)          &      -1.02(5)    &  -1.03  \\[1.5mm] 
Acetamide*            & $2_2 \rightarrow 2_1 A^{\pm}$                 &         9\,254.418(4)          &      -1.04(5)    &  -1.12  \\
                     & $1_0 \rightarrow 1_1 E$                       &        13\,388.703(4)          &      -1.57(8)    &  -1.34  \\
                     & $ 4_3 \rightarrow 4_2 A^{\mp}$                &        14\,210.349(4)          &      -1.03(5)    &  -1.12  \\
                     & $ 3_3 \rightarrow 3_2 A^{\mp} $               &        14\,441.705(4)          &      -1.05(5)    &  -1.12  \\
                     & $2_0 \rightarrow 2_1 E$                       &        15\,115.748(4)          &      -1.43(7)    &  -1.30  \\
                     & $2_2 \rightarrow 1_1 E$                       &        22\,095.527(4)          &      -0.74(5)    &  -0.78  \\
                     & $3_3 \rightarrow 3_2 A^{\pm}$                 &        22\,769.635(4)          &      -1.03(5)    &  -1.08  \\
                     & $4_2 \rightarrow 3_3 E$                       &        47\,373.320(4)          &      -1.04(5)    &  -1.11  \\[1.5mm]
Methyl Formate*       & $1_1 \rightarrow 1_{-1} E$                    &         1\,610.900(2)          &      -1.00(5)    &  -0.70  \\
                     & $2_{-1} \rightarrow 1_{-1} E$                 &        22\,827.741(8)          &      -1.00(5)    &  -1.00  \\
                     & $2_1 \rightarrow 1_1 A^+$                     &        22\,828.134(8)          &      -1.00(5)    &  -1.00  \\
                     & $2_0 \rightarrow 1_0 E$                       &        24\,296.491(8)          &      -1.00(5)    &  -1.00  \\
                     & $2_0 \rightarrow 1_1 A^+$                     &        24\,298.481(8)          &      -1.00(5)    &  -1.00  \\
                     & $2_1 \rightarrow 1_1 E$                       &        26\,044.796(8)          &      -1.00(5)    &  -1.00  \\
                     & $2_1 \rightarrow 1_1 A^-$                     &        26\,048.534(8)          &      -1.00(5)    &  -1.00  \\
                     & $4_0 \rightarrow 3_0 E$                       &        47\,534.069(16)         &      -1.00(5)    &  -1.00  \\[1.5mm]
Acetic Acid*          & $8_{-1} \rightarrow 7_{-1} E$                 &        90\,203.444(20)         &      -1.00(5)    &  -1.00  \\
                     & $8_{-1} \rightarrow 7_0 E$                    &        90\,203.444(20)         &      -1.00(5)    &  -0.97  \\
                     & $8_0 \rightarrow 7_{-1} E$                    &        90\,203.444(20)         &      -1.00(5)    &  -1.03  \\
                     & $8_0 \rightarrow 7_0 E$                       &        90\,203.444(20)         &      -1.00(5)    &  -1.00  \\
                     & $8_0 \rightarrow 7_0 A^+$                     &        90\,246.250(50)         &      -1.00(5)    &  -1.00  \\
                     & $8_0 \rightarrow 7_1 A^+$                     &        90\,246.250(50)         &      -1.00(5)    &  -1.00  \\
                     & $8_1 \rightarrow 7_0 A^+$                     &        90\,246.250(50)         &      -1.00(5)    &  -1.00  \\
                     & $8_1 \rightarrow 7_1 A^+$                     &        90\,246.250(50)         &      -1.00(5)    &  -1.00  \\
    
\\[-1.8ex]

\toprule\\[-1.8ex]
\end{tabular*}
\end{table*}
\endgroup

Using the recipe described above, the sensitivity coefficient of any desired transition in a molecule containing a $C_{3v}$ symmetry group can be calculated. We have calculated the sensitivity coefficients of many ($>1000$) transitions in methanol, acetaldehyde, acetamide, methyl formate and acetic acid, using the constants listed in Refs.~\cite{Xu2008o,Kleiner1996,Ilyushin2004,Carvajal2007,Ilyushin2008}, respectively. In Table~\ref{tab:sensitivities} sensitivity coefficients of selected transitions in the vibrational ground state ($\nu_t=0$) of these molecules are listed. For methanol, the most sensitive transitions involving levels with $J\leq 10$ are listed. For the other molecules no large sensitivities were found for the vibrational ground state and the eight lowest transition frequencies that have been observed in the interstellar medium have been listed~\cite{Lovas2004}. From the table, we see in particular that transitions in methanol are much more sensitive to a variation of $\mu$ than the transitions listed for the other molecules. Except for the $1_{-1}\rightarrow 1_{1}\,E$ transition in acetaldehyde with $K_{\mu}=-3.7$, all transitions in acetaldehyde, acetamide, methyl formate and acetic acid have $-1.57\leq K_{\mu}\leq -0.74$. 

The error in the last digit of the $K_{\mu}$-coefficients is quoted in brackets, and is conservatively taken to be 5\% if $|K_{\mu}| \geq 1$ or 0.05 if $|K_{\mu}| <1$. The error in the sensitivity coefficients has 3 sources: (i) errors due to the uncertainty in the determination of the molecular constants. As the simulations reproduce almost all transitions $<$100~kHz, this error is negligible small. (ii) Errors due to inexactness of the scaling relations of higher order constants. Many of the higher-order constants are products of torsional and rotational operators and may also be fairly correlated. Therefore, the exact relationships between the higher order parameters and the moments of inertia (and masses) are not obvious. For methanol we have investigated the influence of the higher order paramaters by comparing $K_{\mu}$ coefficients calculated by scaling only the first 7 and the first 31 constants to $K_{\mu}$ coefficients obtained by scaling all 119 constants. Sensitivity coefficents were found to typically agree within 5\% or 0.5\% if 7 or 31 constants were scaled, respectively. The effect of the higher order terms is therefore expected to be small. This has been confirmed by an independent study by Levshakov \emph{et al.}~\cite{Levshakov2011}. (iii) Errors due to neglecting the $\mu$ dependence of the torsional potential. Within the Born-Oppenheimer approximation the torsional potential, $V_3$, is independent of the mass of the nuclei and hence of $\mu$. It is known however that $V_3$ does vary between isotopologues. For instance for \ce{^{12}CH3^{16}OH} the torsional potential $V_3\approx 373 \text{\,cm}^{-1}$, and for \ce{^{12}CD3^{16}OD} $V_3\approx 362\text{\,cm}^{-1}$. A reliable model for this variation is not available. As a check, we have assumed $V_3$ to be a linear function of $I_{\text{red}}=I_{a1}I_{a2}/I_{a}$; $V_{3}$=$V_{3}(\ce{^{12}CH3^{16}OH})-19.4 (I_{\text{red}}-I_{\text{red}}(\ce{^{12}CH3^{16}OH}))$. As $I_{\text{red}}$ is directly proportional to $\mu$, this introduces a $\mu$ dependence in the potential. We found that the $K_{\mu}$ coefficients for \ce{^{12}CH3^{16}OH} calculated by including the linear scaling for $V_{3}$ are typically 3\% smaller than those obtained when the potential is assumed to be independent of $\mu$.

\begingroup
\squeezetable
\centering
\renewcommand{\arraystretch}{1.25}
\begin{table*}[tb!]
\caption{Rotational and torsional energy differences and associated sensitivity coefficients. Expressions for differences in rotational energy are only valid in the limit of a symmetric top molecule.}
\label{tab:toymodel}
\begin{tabular*}{1 \textwidth}{@{\extracolsep{\fill}}l l l l l}

\hline\hline
\bigstrut & & &  \multicolumn{1}{c}{$\Delta E$} & \multicolumn{1}{c}{$K_{\mu}$} \\
\hline
\\[-1.8ex]
Rotation & $\Delta{J}=0$&$\Delta{K}=\pm 1$    & $\left [A - \frac{1}{2}(B+C) \right ]\left ( 1 \pm 2K \right )$ & -1\\
                         & $\Delta{J}=\pm 1$&$\Delta{K}=0$    & $\pm {(B+C)} \left [ J + \frac{1}{2}(1\pm 1) \right ]         $& -1\\
                         & $\Delta{J}=\pm 1$&$\Delta{K}=\pm 1$& $\pm {(B+C)} \left [ J + \frac{1}{2}(1\pm 1) \right ] +  \left [A - \frac{1}{2}(B+C) \right ]\left ( 1 \pm 2K \right )$    &  -1 \\
                         & $\Delta{J}=\pm 1$&$\Delta{K}=\mp 1$&  $\pm {(B+C)} \left [ J + \frac{1}{2}(1\pm 1) \right ] +  \left [A - \frac{1}{2}(B+C) \right ]\left ( 1 \mp 2K \right )$                                                &  -1       \\
Torsion &                  &$\Delta{K}=0$\footnote[1]{Note that the transitions involving different overall symmetry species of the torsional levels ($A \leftrightarrow E$) are forbidden, thus transitions with $\Delta K=0$, $\Delta J=0$ do not occur}    &   $\pm\sqrt{3} Fa_1 \sin{\left ( \tfrac{2\pi}{3}\left [\rho K \mp \tfrac{1}{2} \right ] \right )}$                                                 &     $(B_1-1)-\frac{1}{2}C_1\sqrt{s}$   \\
                         &                  &$\Delta{K}=\pm1$    &     $\mp 2Fa_1\sin \left (\frac{\pi}{3}\rho \right ) \sin \left [\frac{2\pi}{3} \left \{ \rho \left ( K \pm \frac{1}{2}\right ) - \sigma \right \}  \right ]$   &  $(B_1-1)-\frac{1}{2}C_1\sqrt{s}$                                                \\
\\[-1.8ex]
\hline\hline
\end{tabular*}
\end{table*}
\endgroup

\section{``Toy" Model\label{sec:toymodel}}

Although the numerical calculations described in the previous section yield the sensitivity coefficient for any desired transition, they provide limited insight. In this section we will devise a simple model which provides an intuitive picture of the physics involved and aids in the identification of other internal rotor molecules that possibly exhibit large sensitivity coefficients. In this model, we neglect coupling between vibrational, rotational and torsional motion. In this case, the $\mu$ dependence of the energy of a certain state $\ket{\nu_{t},J,K,\textit{Ts}}$ can be written as  

\begin{equation}
\mu \left( {\frac{\partial E}{\partial \mu}} \right)_{\nu_{t},J,K,\textit{Ts}}=
K_{\mu}^{\text{vib}} E_{\text{vib}} + K_{\mu}^{\text{rot}} E_{\text{rot}} + K_{\mu}^{\text{tors}} E_{\text{tors}},
\label{eq:leveldependence}
\end{equation}

\noindent
where $K_{\mu}^{\text{vib}}$, $K_{\mu}^{\text{rot}}$, and $K_{\mu}^{\text{tors}}$ are the sensitivities to a possible variation of the proton-to-electron mass ratio of a vibrational, rotational, and torsional transition, respectively. We neglect vibrational excitation and use Eq.~\eqref{eq:leveldependence} to rewrite Eq.~\eqref{eq:K_resonance} as

\begin{equation}
K_{\mu}=\frac{K_{\mu}^{\text{rot}}\Delta{E_{\text{rot}}}+K_{\mu}^{\text{tors}}\Delta{E_{\text{tors}}}}{\Delta{E_{\text{rot}}}+ \Delta{E_{\text{tors}}}},
\label{eq:K_toy_model}
\end{equation}

\noindent
with $\Delta{E_{\text{rot}}}$ and $\Delta{E_{\text{tors}}}$ being the difference in rotational and torsional energy between the two energy levels involved, respectively. From this equation it immediately follows that $K_{\mu}$ diverges for $\Delta{E_{\text{rot}}}=-\Delta{E_{\text{tors}}}$ and $K_{\mu}^{\text{rot}}\neq K_{\mu}^{\text{tors}}$. This implies that the highest sensitivities are expected for transitions that convert overall rotation into internal rotation or vice versa. Furthermore, Eq.~\eqref{eq:K_toy_model} indicates that the $K_{\mu}$ coefficients are proportional to the amount of energy that is cancelled. We will now derive approximate analytical expressions for $\Delta{E_{\text{rot}}}$,
$\Delta{E_{\text{tors}}}$, $K_{\mu}^{\text{rot}}$ and $K_{\mu}^{\text{tors}}$ using the results 
of Sec.~\ref{sec:internalrotation}. 

The rotational energy is given by Eq.~\eqref{eq:rotEn}, from which it is straightforward to calculate the
energy differences for different transitions. The results are listed in Table~\ref{tab:toymodel}.
Note that these expressions are valid only for a nearly symmetric molecule. 

$K_{\mu}^{\text{rot}}$ follows from the $\mu$ dependence of the rotational constants. 
From Eq.~\eqref{eq:A}-\eqref{eq:C} $A$, $B$, and $C$ are inversely proportional
to the moments of inertia that are proportional to $\mu$. Consequently, in first order approximation 
$K_{\mu}^{\text{rot}}=-1$.

\begin{figure}[tb]
\includegraphics[width=\columnwidth]{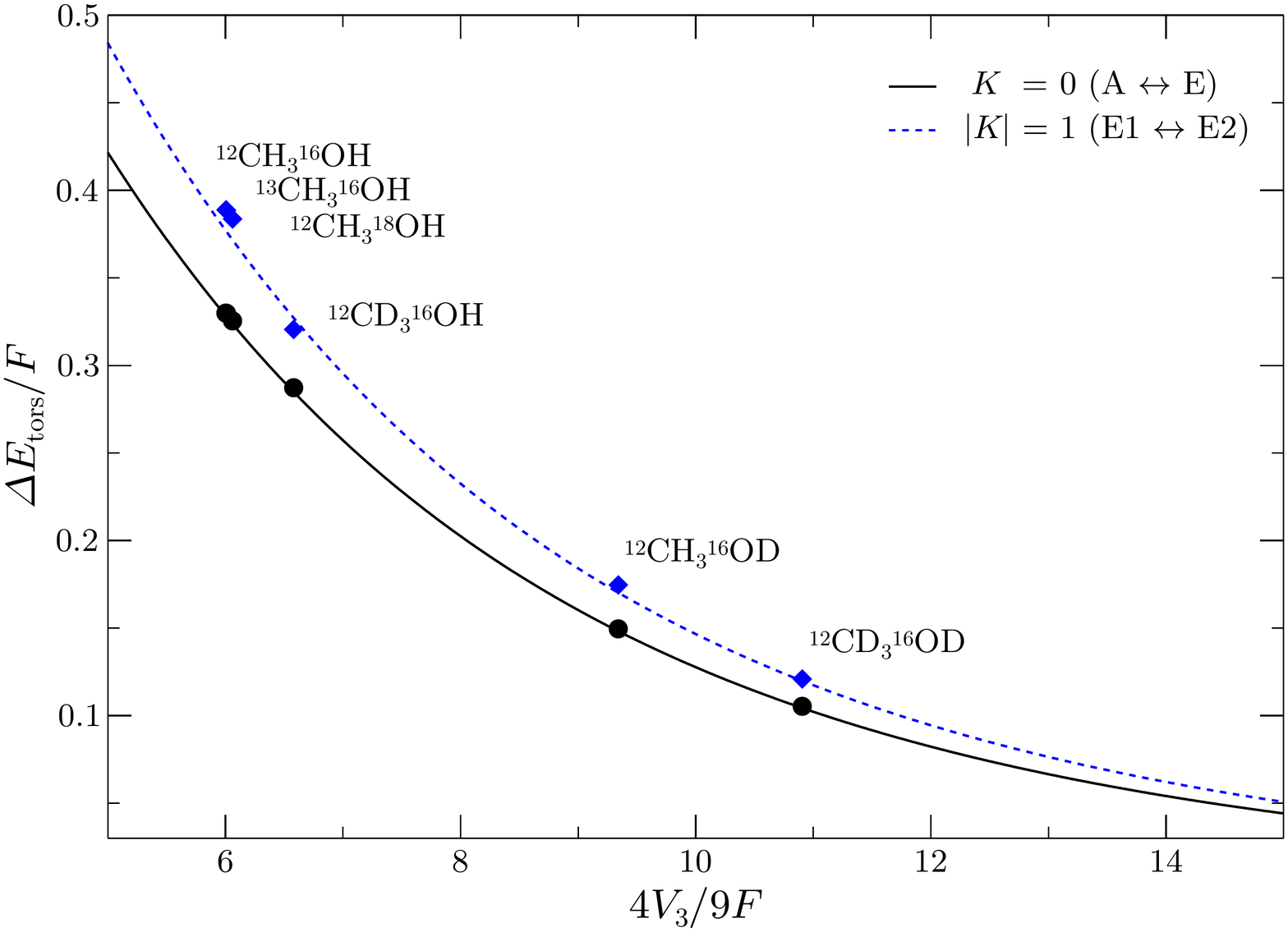}
\caption{Torsional energy splitting between $A$ and $E$ $J = 0, K = 0$ (solid line) and $E1$ and $E2$ $J = 1, |K| = 1$ levels (dashed line) as function of $s$ from Eq.~\eqref{eq:AEsplitting} and~\eqref{eq:EEsplitting}. For the curve representing the splitting between the $E1$ and $E2$ states, $\rho$ is fixed at $\rho=0.8$. Also shown are the torsional splittings for the different isotopologues of methanol. 
\label{fig:isotopologues}}
\end{figure}

The torsional energy is given by Eq.~\eqref{eq:Etors}, from which the splitting between the $A$ $(\sigma=0)$ and $E$ $(\sigma=\pm 1)$ states can be calculated
\begin{align}
\Delta{E_{\text{tors}}^{A\leftrightarrow E}} &= Fa_1 \left [ \cos{\left \{ \tfrac{2\pi}{3} \left ( \rho K \pm 1 \right ) \right \}}  - \cos{\left ( \tfrac{2\pi}{3}\rho K \right )}  \right ]\notag\\
&= \pm\sqrt{3} Fa_1 \sin{\left ( \tfrac{2\pi}{3}\left [\rho K \pm \tfrac{1}{2} \right ] \right )}.\label{eq:AEsplitting}
\end{align}
Analogously, the splitting between $E1$ $(\sigma=+1)$ and $E2$ $(\sigma=-1)$ levels is given by
\begin{equation}
\Delta{E_{\text{tors}}^{E1\leftrightarrow E2}} = \sqrt{3} Fa_1 \sin{\left ( \tfrac{2\pi}{3}\rho K \right )}.\label{eq:EEsplitting}
\end{equation}

Equations~\eqref{eq:AEsplitting}~and~\eqref{eq:EEsplitting} are plotted in Fig.~\ref{fig:isotopologues} as function of $s$ for $K=0$ and $|K|=1$, respectively, using the $A_1$, $B_1$ and $C_1$ parameters obtained in Sec.~\ref{sec:internalrotation}. Also shown are data points for the various isotopologues of methanol derived from experiments. Within the Born-Oppenheimer approximation $s$ is proportional to $I_{\text{red}}\simeq\frac{1}{2}\hbar^2 F^{-1}$.

As transitions between different symmetry states ($A \leftrightarrow E$) are not allowed, we are interested in transitions within the same torsional states that differ in $K$. The difference in torsional energy for such transition with $\Delta K=\pm 1$ can be derived from Eq.~\eqref{eq:Etors} to be
\begin{equation}
\Delta{E_{\text{tors}}} = \mp 2Fa_1\sin{\left ( \tfrac{\pi}{3}\rho \right )}\sin{\left ( \tfrac{2\pi}{3}\left \{\rho \left ( K \pm
\tfrac{1}{2} \right ) +\sigma \right \} \right )}.
\label{eq:KK+1splitting}
\end{equation}

Due to the  $a_1$-coefficient appearing in Eq.~\eqref{eq:AEsplitting}-\eqref{eq:KK+1splitting} all torsional splittings have the same dependence on $s$, and hence on $\mu$. The sensitivity of a torsional transition, $K_{\mu}^{\text{tors}}$, can be obtained from Eq.~\eqref{eq:Kmu} by writing
\begin{align}
K_{\mu}^{\text{tors}} &= \left ( \frac{\partial\Delta{E_{\text{tors}}}}{\partial s} \right ) \left ( \frac{\partial s}{\partial\mu} \right ) \frac{\mu}{\Delta{E_{\text{tors}}}}\notag\\
&=\frac{\partial \left ( \Delta{E_{\text{tors}}}/F \right )}{\partial s} \frac{sF}{\Delta{E_{\text{tors}}}}-1,\label{eq:Kmutors}
\end{align}

\noindent
where we have used the fact that $F$ scales with $\mu^{-1}$. The $-1$ appearing in the second line is introduced by the substitution of $\Delta{E}_{\text{tors}}$ with $\Delta{E}_{\text{tors}}/F$. By inserting Eq.~\eqref{eq:KK+1splitting} in Eq.~\eqref{eq:Kmutors} we obtain

\begin{align}
K_{\mu}^{\text{tors}}&=(B_1-1)-\tfrac{1}{2}C_1\sqrt{s} \notag\\
                     &\simeq 0.111-1.060\sqrt{s},
\label{eq:toy_modelKmu}
\end{align}
\noindent
where we have used the dimensionless fit values for $B_{1}$ and $C_{1}$ obtained in Sec.~\ref{sec:internalrotation}. Hence, within our approximations, $K_{\mu}^{\text{tors}}$ is only a function of $s$. For \ce{^{12}CH3^{16}OH}, with $s=6.01$, this results in $K_{\mu}^{\text{tors}}=-2.5$. In agreement with the value found in Jansen \emph{et al.}~\cite{Jansen2011}. 

\begin{figure}[bt]
\includegraphics[width=\columnwidth]{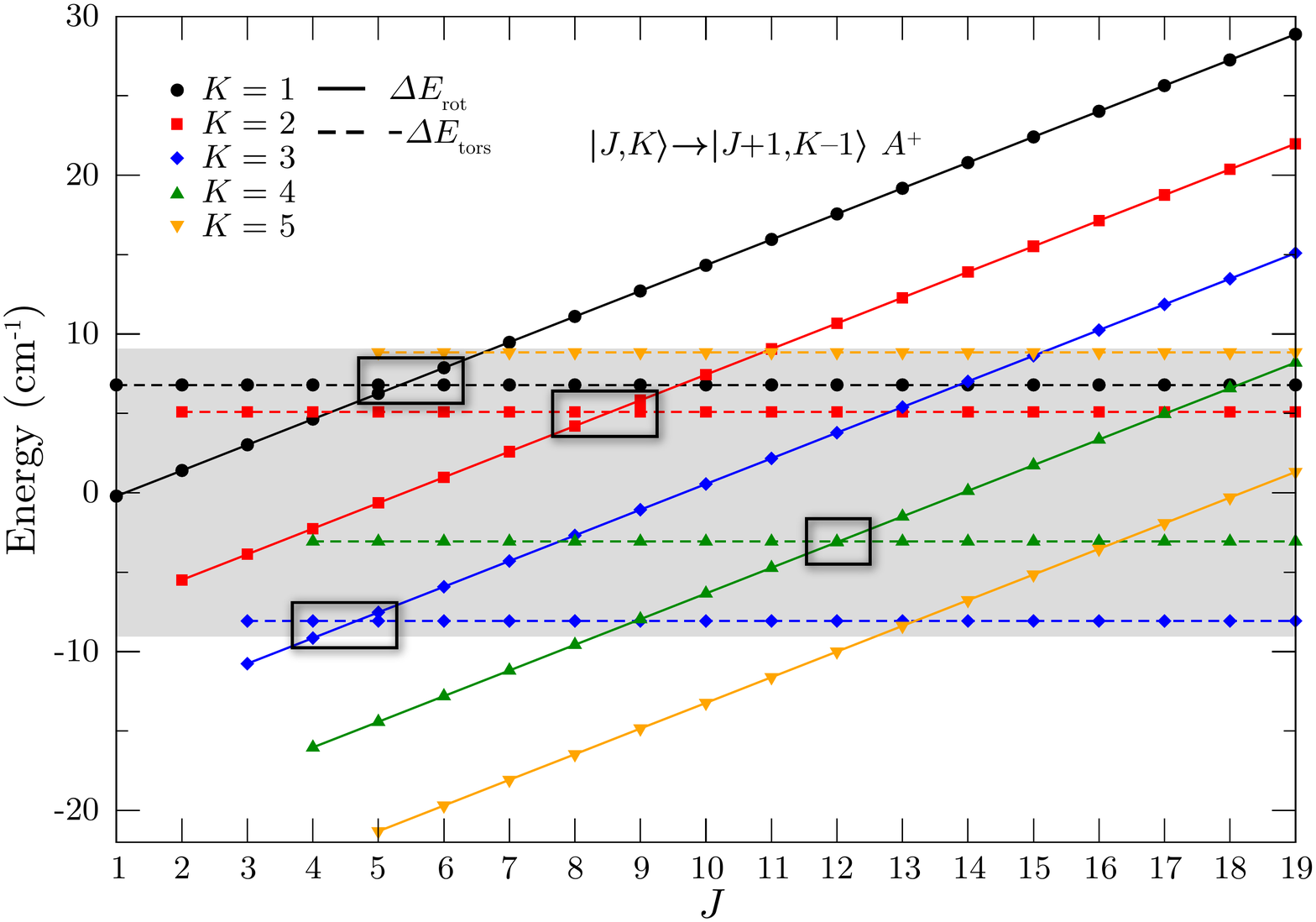}
\caption{Difference in rotational energy, $\Delta E_{\text{rot}}$, (solid lines) and negated difference in torsional energy, $-\Delta E_{\text{tors}}$, (dashed lines) between levels 
connected by a $\ket{J,K}\rightarrow\ket{J+1,K-1}\,A^+$ type transition in methanol. The curves are obtained by the expressions in Table~\ref{tab:toymodel} and the molecular constants of methanol~\cite{Xu2008o}. The highest sensitivities are expected when torsional energy is converted into rotational energy or vice versa -- 
i.e. when the two curves cross. The shaded area represents the maximum torsional energy that can be attained by the molecule. Note that the amount of energy that is cancelled is proportional to the $K_{\mu}$ coefficient. 
\label{fig:deltaErotEtors}}
\end{figure}

With the help of Eq.~\eqref{eq:K_toy_model} and the expressions for $\Delta{E_{\text{rot}}}$, 
$\Delta{E_{\text{tors}}}$, $K_{\mu}^{\text{rot}}$ and $K_{\mu}^{\text{tors}}$ as listed in Table~\ref{tab:toymodel},
we can now determine which transitions are likely to have an enhanced sensitivity and
estimate the $K_{\mu}$ coefficients of these transitions. From Eq.~\eqref{eq:K_toy_model}, 
we saw that the highest sensitivities are expected when  
$\Delta{E_{\text{rot}}}\simeq -\Delta{E_{\text{tors}}}$. In Fig.~\ref{fig:deltaErotEtors} 
the difference in rotational energy, $\Delta{E_{\text{rot}}}$, (solid curves) 
and negated torsional energy, $-\Delta{E_{\text{rot}}}$ (dashed curves) are shown for 
$\ket{J,K}\rightarrow\ket{J+1,K-1}A^+$ transitions in methanol. The highest sensitivities are expected 
when the lines representing the difference in rotational energy $\Delta{E_{\text{rot}}}$ and the negated difference in torsional energy $-\Delta{E_{\text{tors}}}$ cross. For this to happen at low $J$ and $K$ it requires that the rotational constants $A$, $B$ and $C$ are of the same order as the difference in torsional energy. If the rotational constants are much smaller or larger than $\Delta{E_{\text{tors}}}$, the crossings will only occur for high $J$ and/or $K$ quantum numbers. With the help of Fig.~\ref{fig:deltaErotEtors}, it is straightforward to select transitions that are likely to have a large $K_{\mu}$ coefficient. For instance, the solid line representing the difference in rotational energy for $K=1$ (solid circles), crosses the dashed line representing the negated difference in torsional energy near $J=5$. As the lines cross near the maximum of torsional energy that may be attained, represented by the border of the gray shaded area, we may expect a large $K_{\mu}$ for the $5_1\rightarrow 6_0\,A^+$ transition. Indeed, this transition has $K_{\mu}=-42$. 

\begin{figure}[bt]
\includegraphics[width=\columnwidth]{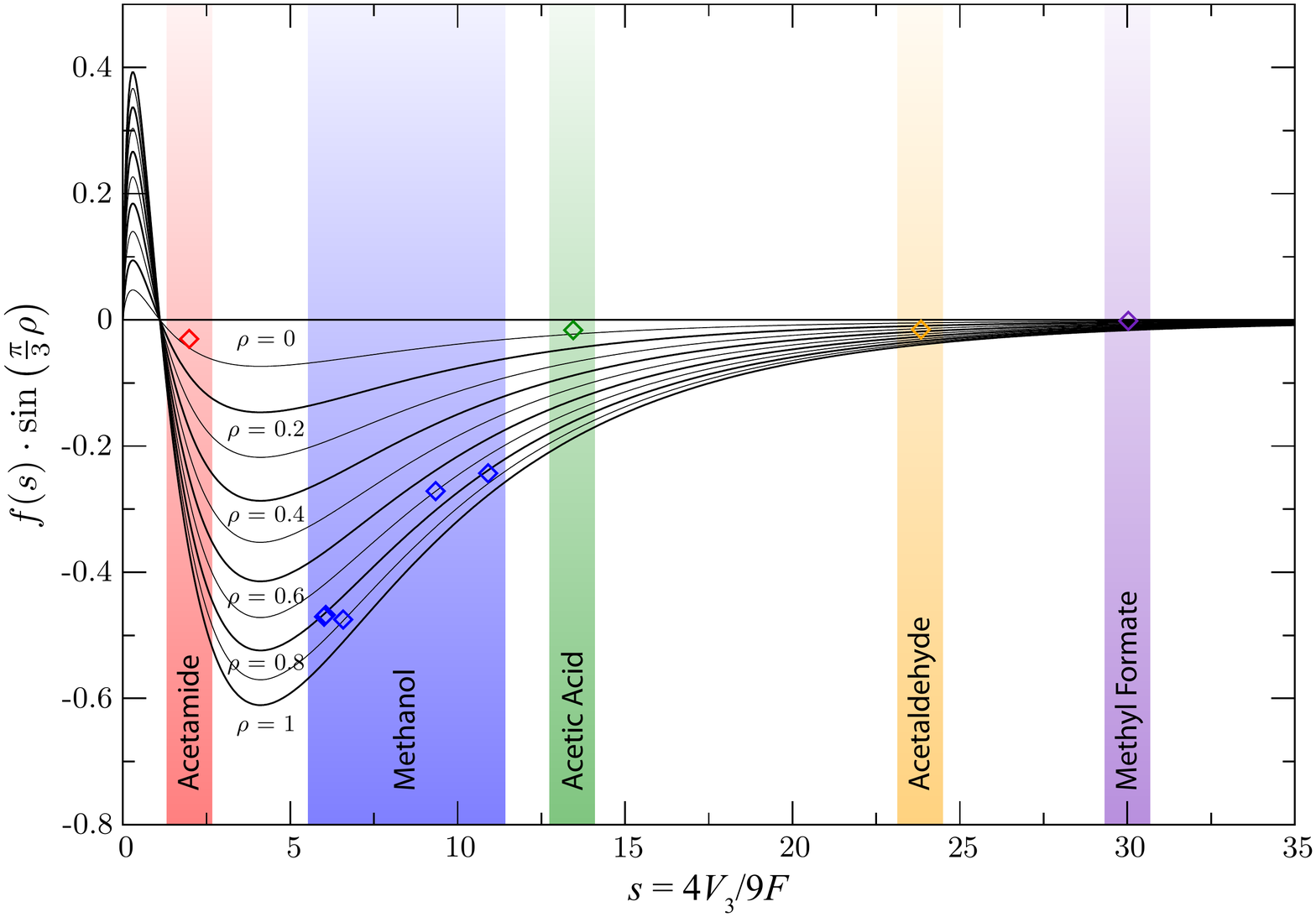}
\caption{The product $f(s)\cdot\sin{\left (\tfrac{\pi}{3}\rho \right )}$ (see text) in the ground torsional state as a function of the effective barrier height, $s$, using Eq.~\eqref{eq:K_toy_model2} with $\rho$ as indicated in the figure. Also shown are data points for each molecule investigated in this paper, values of $s$ and $\rho$ can be found in Table~\ref{tab:molecules&constants}.
\label{fig:Kmu-toy}}
\end{figure}

The sensitivity coefficients $K_{\mu}$ of the transitions can be estimated using Eq.~\eqref{eq:K_toy_model}. Unfortunately, we found that the agreement between the $K_{\mu}$ coefficients obtained from this simple model and the values found from the full calculation was unsatisfactory, mainly as a result of neglecting the assymmetry of the molecules. Hence we chose to use the experimental energy difference between the levels, $h \nu$, rather than $(\Delta{E_{\text{rot}}}+ \Delta{E_{\text{tors}}})$. In this case Eq.~\eqref{eq:K_toy_model} can be written as

\begin{align}
K_{\mu} &= \left( K_{\mu}^{\text{rot}}\left(h\nu-\Delta{E_{\text{tors}}}\right)
+ K_{\mu}^{\text{tors}}\Delta{E_{\text{tors}}}\right) / h \nu \notag\\
&= K_{\mu}^{\text{rot}}+\Delta{E_{\text{tors}}}\left(K_{\mu}^{\text{tors}}-K_{\mu}^{\text{rot}}\right)/ h \nu\notag\\
&=-1 \mp 2F A_1 s^{B_1} e^{-C_1\sqrt{s}} \left( B_1- \tfrac{1}{2}C_1\sqrt{s} \right) \times \notag\\
&~~~~~~~~\sin{\left ( \tfrac{\pi}{3}\rho \right )}\sin{\left( \tfrac{2\pi}{3}\left \{\rho \left ( K \pm \tfrac{1}{2} \right )
+\sigma \right \} \right)}/h \nu \notag\\
&\equiv -1 + F \cdot f(s) \cdot \sin{\left ( \tfrac{\pi}{3}\rho \right )} \cdot g(\rho,K,\sigma)/ h\nu, 
\label{eq:K_toy_model2}
\end{align}

\noindent
The calculated $K_{\mu}$ coefficients from this model for selected transitions are listed in Table~\ref{tab:sensitivities} and are seen to agree rather well with the numerical calculations.  

\begingroup
\squeezetable
\centering
\begin{table*}[tb!]
\caption{Structure and some lower-order constants of the molecules investigated in this paper. The $A-E$ $K=0$ torsional splitting from Eq.~\eqref{eq:AEsplitting} is listed for the vibrational ground state $\nu_t=0$ of these molecules. Note that the magnitude of the torsional splitting only depends on the reduced barrier height $s=4V_3/9F$. The magnitude of $K_{\mu}^{\text{tors}}$ for each molecule follows from Eq.~\eqref{eq:Kmutors}. The $K_{\mu}^{\text{gen}}$ values given in the last column of the table are hypothetical sensitivities that may be expected for transitions of 1\,GHz and with $g(\rho,K,\sigma)=1$. Molecular constants are taken from Refs.~\cite{Xu2008o,Kleiner1996,Ilyushin2004,Carvajal2007,Ilyushin2008}.}
\label{tab:molecules&constants}
\begin{tabular}{l l l D..{5} D..{5} D..{5} D..{5} D..{5} D..{5} D..{5}}
\hline\hline
& \multicolumn{1}{l}{Structure}
& \multicolumn{1}{l}{Isotopologue}
& \multicolumn{1}{c}{$V_3$}
& \multicolumn{1}{c}{$F$}
& \multicolumn{1}{c}{$s$}
& \multicolumn{1}{c}{$\rho$}
& \multicolumn{1}{c}{$\Delta E_{\text{tors}}^{A\leftrightarrow E}$}
& \multicolumn{1}{c}{$K_{\mu}^{\text{tors}}$}
& \multicolumn{1}{c}{$K_{\mu}^{\text{gen}}+1$ \bigstrut} \\
& & & \multicolumn{1}{c}{(cm$^{-1}$)} &  \multicolumn{1}{c}{(cm$^{-1}$)} & & &  \multicolumn{1}{c}{(cm$^{-1}$)}&&\bigstrut \\
\hline
\\[-1.8ex]

\multirow{6}{*}{Methanol}& \multirow{6}{*}{{\tiny\chemfig{[:90]C(-[::-90]O(-[::72]H))(-[::-225]H)(<[::-290]H)(<:[::-340]H)}\newline}}& \ce{^{12}CH3^{16}OH} & 373.555 &  27.647 & 6.01 & 0.810 & 9.07 & -2.5 & \pm 398 \\[0.3ex]
                             &                                              & \ce{^{13}CH3^{16}OH} & 373.777 & 27.642  & 6.01 & 0.810 & 9.06 & -2.5 & \pm 397 \\[0.3ex]
                             &                                              & \ce{^{12}CH3^{18}OH} & 374.067 & 27.428  & 6.06 & 0.809 & 8.88 & -2.5 & \pm 392 \\[0.3ex]
                             &                                              & \ce{^{12}CD3^{16}OH} & 370.055 & 24.994  & 6.58 & 0.895 & 7.12 & -2.6 & \pm 363 \\[0.3ex]
                             &                                              & \ce{^{12}CH3^{16}OD} & 366.340 & 17.428  & 9.34 & 0.699 & 2.58 & -3.1 & \pm 145 \\[0.3ex]
                             &                                              & \ce{^{12}CD3^{16}OD} & 362.122 & 14.758  &10.91 & 0.822 & 1.54 & -3.4 & \pm 110 \\

Acetaldehyde \rule{0pt}{14 mm}& {\tiny\chemfig{[:90]C(-[::-90]C(=[::-57]O)(-[::63]H))(-[::-225]H)(<[::-290]H)(<:[::-340]H)}\newline} & \ce{^{12}CH3^{12}C^{16}OH} & 407.716 & 7.600 & 23.84 & 0.332 & 0.065 & -5.1 & \pm 3.6 \\
\\[-1.8ex]
Acetamide  \rule{0pt}{14 mm}& {\tiny\chemfig{[:90]C(-[::-90]C(-[::60]N(-[::60]H)(-[::-60]H))(=[::-60]O))(-[::-225]H)(<[::-290]H)(<:[::-340]H)}\newline} &\ce{^{12}CH3^{12}C^{16}O^{14}NH2}& 25.044 & 5.617 & 1.98 & 0.068 & 4.85 & -1.4 & \pm 5.2 \\
\\[-1.8ex]
Methyl Formate  \rule{0pt}{14 mm}& {\tiny\chemfig{[:90]C(-[::-90]O(-[::66]C(=[::55]O)(-[::-71]H)))(-[::-225]H)(<[::-290]H)(<:[::-340]H)}\newline} & \ce{^{12}CH3^{16}O^{12}C^{16}OH}& 370.924 & 5.490 & 30.03 & 0.084 & 0.017 & -5.7 & \pm 0.3 \\
\\[-1.8ex]
Acetic Acid \rule{0pt}{14 mm}& {\tiny\chemfig{[:90]C(-[::-90]C(=[::-54]O)(-[::68]O(-[::-74]H)))(-[::-225]H)(<[::-290]H)(<:[::-340]H)}\newline} & \ce{^{12}CH3^{12}C^{16}O^{16}OH} & 170.174 & 5.622 & 13.45 & 0.072 & 0.34 & -3.8 & \pm 2.8 \\
\\[-1.8ex] 

\hline\hline

\end{tabular}

\end{table*}
\endgroup

We are now ready for a qualitative discussion of the sensitivity coefficients obtained for the different molecules. In the last line of Eq.~\eqref{eq:K_toy_model2}, we have separated the expression for the sensitivity coefficient in four parts. The molecular constant $F$, a function $f(s)$ that depends only on $s$, a function that depends only on $\rho$ and a function, $g(\rho,K,\sigma)$ that depends on the rotational quantum number, $K$, on the torsional symmetry, $\sigma$, and on $\rho$. This last function takes on a value between -1 and 1 for the different $\sigma$, $K$ levels. Although this function determines the sensitivity of a specific level, it is not important for comparing different molecules. The product $f(s)\cdot\sin{\left (\tfrac{\pi}{3}\rho \right )}$ can be used as a means to compare the sensitivity for different molecules. In Fig.~\ref{fig:Kmu-toy} this product is plotted as a function of $s$ with $\rho$ as indicated in the figure. The curves can be regarded as the maximum sensitivity one may hope to find in a molecule with a certain $F$ and transition energy $h\nu$, if $g(\rho,K,\sigma)$ is set to $\pm 1$. The maximum sensitivity peaks at $s=4$ and $\rho=1$. Recall that $\rho$ is defined as the moment of inertia of the top over the moment of inertia of the whole molecule ($\rho\simeq I_{a2}/I_a$), and cannot be greater than unity. 

In Table~\ref{tab:molecules&constants} the structure and lowest order molecular constants of the molecules investigated in this paper are listed as well as the results from our analytical model. The last column lists the generic maximum sensitivity $K_{\mu}^{\text{gen}}$ that may be expected for a hypothetical transition with a frequency of 1\,GHz. For methanol, this number is 10--100 times larger than for the other investigated molecules, following from the fact that methanol has a large $F$, an effective barrier close to the optimal value of 4, and a relatively large $\rho$. 

\subsection*{Application of the Toy model to excited torsional states}
So far, we have limited the discussion to transitions in the ground torsional state ($\nu_t=0$). In this paragraph we discuss the application of the model to excited torsional states. Transitions within excited torsional states are both unlikely to be observed in the interstellar medium and of less interest for laboratory tests due to the added complexity of the experiment, hence, the following discussion is intended for providing a more complete picture only. In excited torsional states the splitting between the different torsional symmetry levels becomes larger as tunneling through the torsional potential becomes more likely. This implies that the energy that can be cancelled also becomes larger, but, at the same time, the sensitivity coefficient of a pure torsional transition, $K_{\mu}^{\text{tors}}$, becomes smaller. As we will see, the first effect is more important.

In Fig.~\ref{fig:expansion_coefficients} the solid squares represent $a_1$ coefficients for the $\nu_t=1$ state obtained by fitting the second eigenvalues of the torsional Hamiltonian as function of $K$ to the Fourier expansion of Eq.~\eqref{eq:Etors}. The dashed line in Fig.~\ref{fig:expansion_coefficients} represents a fit using an expression similar to Eq.~\eqref{eq:a1}
with an additional term, i.e.,

\begin{equation}
a_1^{\nu_t=1} = A_1{s^{B_1}}\cdot e^{-C_1\sqrt{s} +D_1 s},\label{eq:a1vt=1}
\end{equation} 
\noindent
with $A_1=10.388$, $B_1=0.829$, $C_1=1.108$ and $D_1=-0.058$. The open squares, also shown in Fig.~\ref{fig:expansion_coefficients}, are $a_1$ coefficients for the first excited torsional state of the five molecules investigated in this paper. The additional term modifies Eq.~\eqref{eq:K_toy_model2} only slightly. 
With the known coefficients, $A_1$, $B_1$, $C_1$ and $D_1$, we can again plot the parts that depend only on $s$ and $\rho$, as was done for $\nu_t=0$ shown in Fig.~\ref{fig:Kmu-toy}. As compared to the $\nu_t=0$, the curves for the $\nu_t=1$ are broadened and the center of the peak is shifted from $s=4$ to $s=7$. Moreover, the generic sensitivity is $5$ times larger at the peak. As a consequence, the generic sensitivity, i.e., the sensitivity for a transition with $\nu=1$\,GHz and $g(\rho,K,\sigma)=1$, of molecules with unfavorable $s$ and $\rho$ in the ground torsional state, can be large in the $\nu_t=1$ state. For instance, acetaldehyde in the $\nu_t=1$ state has a generic sensitivity $K_{\mu}^{\text{gen}}+1=\pm 82$ as compared to $K_{\mu}^{\text{gen}}+1=\pm 3.6$  in the torsional ground state. For methanol, the generic sensitivity will also increase, however as the torsional splitting in the $\nu_t=1$ state is much larger than the rotational splittings, resonances are expected to occur only at high $J$ and $K$ values. Note that the sensitivities from transitions from the $\nu_t=0$ to the $\nu_t=1$ state will not be significantly enhanced as compared to ordinary vibrational transitions, i.e. $K_{\mu}\approx-0.5$.  

\section{Conclusion \& Outlook}

In the present study we have demonstrated that transitions in internal rotor molecules that convert internal rotation energy into overall rotation energy of the molecule exhibit an enhanced sensitivity to a possible variation of the proton-to-electron mass ratio. We have calculated the $K_{\mu}$ coefficients for five, relatively small, internal rotor molecules that are of astrophysical relevance; methanol, acetaldehyde, acetamide, methyl formate and acetic acid. In addition to full calculations using advanced codes such as the {\sc belgi} program achieving spectroscopic accuracy on the level energies, we have developed an approximate model, dubbed as ``toy"-model, in which the molecular structure is described by the six most relevant of the molecular parameters. Based on this model we produce insight in the question as to why certain molecules of $C_{3v}$ symmetry with hindered internal rotation are sensitive to $\mu$-variation. In particular, molecules in the torsional ground state are expected to have large $K_{\mu}$ coefficients if they have (i) an effective barrier height, $s$, around 4, (ii) a ratio between the moment of inertia of the top and the whole molecule, $\rho$, close to unity, and (iii) a large value for the molecular constant that relates to the internal rotation, $F$. If the torsional splittings are of the same order as the rotational constants, sensitive transitions will occur between levels with low $J$ and $K$ quantum numbers.

From the approximate ``toy" model we learn that of the five molecules studied, methanol has by far the largest sensitivity, due to its favorable value of the effective barrier $s$, and the fact that in methanol $\rho$ is near unity. Moreover methanol has a fairly large value of $F$. The other investigated molecules either have a too large barrier (acetaldehyde), have a too heavy frame attached to the methyl group and consequently a very small $\rho$ (acetamide and acetic acid), or have both a high barrier and a small $\rho$ (methyl formate). Based on these criteria, other interesting molecules containing a $C_{3v}$ symmetry group include mercaptan, \ce{CH3SH}, ($F=15$ cm$^{-1}$, $V_3 = 439$ cm$^{-1}$ and $\rho = 0.65$, resulting in $K_{\mu}^{\text{gen}}+1=\pm 67$) and methylamine (which will be topic of a separate publication \cite{Ilyushin2011}). Other interesting candidate molecules, although of $C_2$ symmetry, are \ce{H2O2} (recently treated by Kozlov~\cite{Kozlov2011b}) and \ce{H2S2}. These molecules require a modification in the definition of $s$ and will have different $A_1$, $B_1$ and $C_1$ coefficients. 

The high sensitivities of internal rotor molecules, particularly methanol, make them excellent target species to search for a variation of the proton to electron mass ratio over cosmological time scales.
It is important to note that the sensitivity coefficients of the transitions in these molecules have both large positive and large negative values, i.e., if the proton-to-electron mass ratio varies, some transitions will shift to higher frequencies while others shift to lower frequencies. This makes it possible to perform a test of the variation of the proton-to-electron mass ratio using transitions in
pertainig to a single species, thus avoiding the many systematic effects that plague tests based on comparing transitions in different molecules. 

Currently, the most stringent bounds on a cosmological variation of $\mu$ is set by observations of hydrogen molecules in high ($z=2-3$) redshift objects~\cite{Malec2010,vanWeerdenburg2011}
and a comparison between the ammonia tunneling frequencies and rotational transitions in anchor molecules at intermediate redshift $(z=0.5-1)$ objects~\cite{Murphy2008,Henkel2009,Kanekar2011}. Methanol provides a system that should result in more stringent bounds for $\mu$-variation. Recently, methanol, as well as methylamine and acetaldehyde have been observed in the gravitationally lensed system, PKS 1830-211 at $z=0.89$ \cite{muller2011,Henkel2011}. The $1_0 \rightarrow 2_{-1}\,E$ transition in methanol reported by Muller \emph{et al.} was calculated to have $K_{\mu}=-7.4$. We have also calculated the $K_{\mu}$ coefficients for the nine observed transitions in acetaldehyde and found that all lines have sensitivities of $K_{\mu}\simeq-1$. Sensitivity coefficients for methylamine, for which three lines were observed in the same survey, will be calculated in a separate paper~\cite{Ilyushin2011}.

The high sensitivity coefficients in methanol are also beneficial for probing variation of $\mu$ as a result of chameleon-like scalar fields. These fields predict a dependence of $\mu$ on the local matter density. Note that the physical origin of these chameleon like theories are very different from theories describing temporal $\mu$ variation. Levshakov \emph{et al.}~\cite{Levshakov2010} compared ammonia spectra taken at high (terrestrial) and low (interstellar) densities of baryonic matter, and observed a statistically significant variation of $\mu$. Recently, a preliminary tests using methanol was performed by Levshakov \emph{et al.}~\cite{Levshakov2011}.  This test obtained similar results as for ammonia but can be further improved if more accurate laboratory data becomes available.

Methanol is also a promising candidate for laboratory tests on a possible variation of $\mu$. In laboratory experiments rare isotopologues can be used, in contrast to cosmological searches. Hence the most sensitive transitions deriving from the present calculations can be targeted; a combined measurement of the $2_1 \rightarrow 1_1\,E$ and $3_0 \rightarrow 4_{-1}\,E$ lines in \ce{CD3OH} exhibit a sensitivity spread $\Delta K_{\mu}$ that is more than 400 times larger than a pure rotational transition~\cite{Jansen2011}. Note that the large dipole moment and low mass of methanol make it possible to use advanced laboratory techniques for cooling and manipulating molecules by electric fields~\cite{vandeMeerakker2008,Bethlem2008}.

\begin{acknowledgments}
This research has been supported by NWO via a VIDI-grant and by the ERC via a Starting Grant. IK acknowledges the financial support provided by ANR-08-BLAN-0054. LHX gratefully acknowledges financial support from the Natural Science and Engineering Research Council of Canada. We thank Mikhail Kozlov, Jon Hougen and Vadim Ilyushin for helpful discussions. 
\end{acknowledgments}


\begin{thebibliography}{39}%
\makeatletter
\providecommand \@ifxundefined [1]{%
 \@ifx{#1\undefined}
}%
\providecommand \@ifnum [1]{%
 \ifnum #1\expandafter \@firstoftwo
 \else \expandafter \@secondoftwo
 \fi
}%
\providecommand \@ifx [1]{%
 \ifx #1\expandafter \@firstoftwo
 \else \expandafter \@secondoftwo
 \fi
}%
\providecommand \natexlab [1]{#1}%
\providecommand \enquote  [1]{``#1''}%
\providecommand \bibnamefont  [1]{#1}%
\providecommand \bibfnamefont [1]{#1}%
\providecommand \citenamefont [1]{#1}%
\providecommand \href@noop [0]{\@secondoftwo}%
\providecommand \href [0]{\begingroup \@sanitize@url \@href}%
\providecommand \@href[1]{\@@startlink{#1}\@@href}%
\providecommand \@@href[1]{\endgroup#1\@@endlink}%
\providecommand \@sanitize@url [0]{\catcode `\\12\catcode `\$12\catcode
  `\&12\catcode `\#12\catcode `\^12\catcode `\_12\catcode `\%12\relax}%
\providecommand \@@startlink[1]{}%
\providecommand \@@endlink[0]{}%
\providecommand \url  [0]{\begingroup\@sanitize@url \@url }%
\providecommand \@url [1]{\endgroup\@href {#1}{\urlprefix }}%
\providecommand \urlprefix  [0]{URL }%
\providecommand \Eprint [0]{\href }%
\providecommand \doibase [0]{http://dx.doi.org/}%
\providecommand \selectlanguage [0]{\@gobble}%
\providecommand \bibinfo  [0]{\@secondoftwo}%
\providecommand \bibfield  [0]{\@secondoftwo}%
\providecommand \translation [1]{[#1]}%
\providecommand \BibitemOpen [0]{}%
\providecommand \bibitemStop [0]{}%
\providecommand \bibitemNoStop [0]{.\EOS\space}%
\providecommand \EOS [0]{\spacefactor3000\relax}%
\providecommand \BibitemShut  [1]{\csname bibitem#1\endcsname}%
\let\auto@bib@innerbib\@empty
\bibitem [{\citenamefont {Uzan}(2003)}]{Uzan2003}%
  \BibitemOpen
  \bibfield  {author} {\bibinfo {author} {\bibfnamefont {J.-P.}\ \bibnamefont
  {Uzan}},\ }\href {\doibase 10.1103/RevModPhys.75.403} {\bibfield  {journal}
  {\bibinfo  {journal} {Rev. Mod. Phys.}\ }\textbf {\bibinfo {volume} {75}},\
  \bibinfo {pages} {403} (\bibinfo {year} {2003})},\ \bibinfo {note} {and
  references therein}\BibitemShut {NoStop}%
\bibitem [{\citenamefont {Varshalovich}\ and\ \citenamefont
  {Levshakov}(1993)}]{Varshalovich1993}%
  \BibitemOpen
  \bibfield  {author} {\bibinfo {author} {\bibfnamefont {D.}~\bibnamefont
  {Varshalovich}}\ and\ \bibinfo {author} {\bibfnamefont {S.}~\bibnamefont
  {Levshakov}},\ }\href
  {http://www.jetpletters.ac.ru/ps/1187/article_17908.shtml} {\bibfield
  {journal} {\bibinfo  {journal} {JETP Lett.}\ }\textbf {\bibinfo {volume}
  {58}},\ \bibinfo {pages} {231} (\bibinfo {year} {1993})}\BibitemShut
  {NoStop}%
\bibitem [{\citenamefont {van Veldhoven}\ \emph {et~al.}(2004)\citenamefont
  {van Veldhoven}, \citenamefont {K\"{u}pper}, \citenamefont {Bethlem},
  \citenamefont {Sartakov}, \citenamefont {van Roij},\ and\ \citenamefont
  {Meijer}}]{Veldhoven2004}%
  \BibitemOpen
  \bibfield  {author} {\bibinfo {author} {\bibfnamefont {J.}~\bibnamefont {van
  Veldhoven}}, \bibinfo {author} {\bibfnamefont {J.}~\bibnamefont
  {K\"{u}pper}}, \bibinfo {author} {\bibfnamefont {H.~L.}\ \bibnamefont
  {Bethlem}}, \bibinfo {author} {\bibfnamefont {B.}~\bibnamefont {Sartakov}},
  \bibinfo {author} {\bibfnamefont {A.~J.~A.}\ \bibnamefont {van Roij}}, \ and\
  \bibinfo {author} {\bibfnamefont {G.}~\bibnamefont {Meijer}},\ }\href
  {http://dx.doi.org/10.1140/epjd/e2004-00160-9} {\bibfield  {journal}
  {\bibinfo  {journal} {Eur. Phys. J. D}\ }\textbf {\bibinfo {volume} {31}},\
  \bibinfo {pages} {337} (\bibinfo {year} {2004})}\BibitemShut {NoStop}%
\bibitem [{\citenamefont {Flambaum}\ and\ \citenamefont
  {Kozlov}(2007{\natexlab{a}})}]{FlambaumNH3_2007}%
  \BibitemOpen
  \bibfield  {author} {\bibinfo {author} {\bibfnamefont {V.~V.}\ \bibnamefont
  {Flambaum}}\ and\ \bibinfo {author} {\bibfnamefont {M.~G.}\ \bibnamefont
  {Kozlov}},\ }\href {\doibase 10.1103/PhysRevLett.98.240801} {\bibfield
  {journal} {\bibinfo  {journal} {Phys. Rev. Lett.}\ }\textbf {\bibinfo
  {volume} {98}},\ \bibinfo {pages} {240801} (\bibinfo {year}
  {2007}{\natexlab{a}})}\BibitemShut {NoStop}%
\bibitem [{\citenamefont {Kozlov}\ and\ \citenamefont
  {Levshakov}(2011)}]{Kozlov2011}%
  \BibitemOpen
  \bibfield  {author} {\bibinfo {author} {\bibfnamefont {M.~G.}\ \bibnamefont
  {Kozlov}}\ and\ \bibinfo {author} {\bibfnamefont {S.~A.}\ \bibnamefont
  {Levshakov}},\ }\href {http://stacks.iop.org/0004-637X/726/i=2/a=65}
  {\bibfield  {journal} {\bibinfo  {journal} {Astrophys. J.}\ }\textbf
  {\bibinfo {volume} {726}},\ \bibinfo {pages} {65} (\bibinfo {year}
  {2011})}\BibitemShut {NoStop}%
\bibitem [{\citenamefont {Flambaum}(2006)}]{Flambaum2006}%
  \BibitemOpen
  \bibfield  {author} {\bibinfo {author} {\bibfnamefont {V.~V.}\ \bibnamefont
  {Flambaum}},\ }\href {\doibase 10.1103/PhysRevA.73.034101} {\bibfield
  {journal} {\bibinfo  {journal} {Phys. Rev. A}\ }\textbf {\bibinfo {volume}
  {73}},\ \bibinfo {pages} {034101} (\bibinfo {year} {2006})}\BibitemShut
  {NoStop}%
\bibitem [{\citenamefont {Flambaum}\ and\ \citenamefont
  {Kozlov}(2007{\natexlab{b}})}]{Flambaum&Kozlov2007}%
  \BibitemOpen
  \bibfield  {author} {\bibinfo {author} {\bibfnamefont {V.~V.}\ \bibnamefont
  {Flambaum}}\ and\ \bibinfo {author} {\bibfnamefont {M.~G.}\ \bibnamefont
  {Kozlov}},\ }\href {\doibase 10.1103/PhysRevLett.99.150801} {\bibfield
  {journal} {\bibinfo  {journal} {Phys. Rev. Lett.}\ }\textbf {\bibinfo
  {volume} {99}},\ \bibinfo {pages} {150801} (\bibinfo {year}
  {2007}{\natexlab{b}})}\BibitemShut {NoStop}%
\bibitem [{\citenamefont {DeMille~\emph{et al.}}(2008)}]{DeMille2008}%
  \BibitemOpen
  \bibfield  {author} {\bibinfo {author} {\bibfnamefont {D.}~\bibnamefont
  {DeMille~\emph{et al.}}},\ }\href {\doibase 10.1103/PhysRevLett.100.043202}
  {\bibfield  {journal} {\bibinfo  {journal} {Phys. Rev. Lett.}\ }\textbf
  {\bibinfo {volume} {100}},\ \bibinfo {pages} {043202} (\bibinfo {year}
  {2008})}\BibitemShut {NoStop}%
\bibitem [{\citenamefont {Bethlem}\ and\ \citenamefont
  {Ubachs}(2009)}]{Bethlem&Ubachs2008}%
  \BibitemOpen
  \bibfield  {author} {\bibinfo {author} {\bibfnamefont {H.~L.}\ \bibnamefont
  {Bethlem}}\ and\ \bibinfo {author} {\bibfnamefont {W.}~\bibnamefont
  {Ubachs}},\ }\href {\doibase 10.1039/B819099B} {\bibfield  {journal}
  {\bibinfo  {journal} {Farad. Disc.}\ }\textbf {\bibinfo {volume} {142}},\
  \bibinfo {pages} {25} (\bibinfo {year} {2009})}\BibitemShut {NoStop}%
\bibitem [{\citenamefont {Jansen}\ \emph {et~al.}(2011)\citenamefont {Jansen},
  \citenamefont {Xu}, \citenamefont {Kleiner}, \citenamefont {Ubachs},\ and\
  \citenamefont {Bethlem}}]{Jansen2011}%
  \BibitemOpen
  \bibfield  {author} {\bibinfo {author} {\bibfnamefont {P.}~\bibnamefont
  {Jansen}}, \bibinfo {author} {\bibfnamefont {L.-H.}\ \bibnamefont {Xu}},
  \bibinfo {author} {\bibfnamefont {I.}~\bibnamefont {Kleiner}}, \bibinfo
  {author} {\bibfnamefont {W.}~\bibnamefont {Ubachs}}, \ and\ \bibinfo {author}
  {\bibfnamefont {H.~L.}\ \bibnamefont {Bethlem}},\ }\href {\doibase
  10.1103/PhysRevLett.106.100801} {\bibfield  {journal} {\bibinfo  {journal}
  {Phys. Rev. Lett.}\ }\textbf {\bibinfo {volume} {106}},\ \bibinfo {pages}
  {100801} (\bibinfo {year} {2011})}\BibitemShut {NoStop}%
\bibitem [{\citenamefont {Lin}\ and\ \citenamefont
  {Swalen}(1959)}]{Lin&Swalen1959}%
  \BibitemOpen
  \bibfield  {author} {\bibinfo {author} {\bibfnamefont {C.~C.}\ \bibnamefont
  {Lin}}\ and\ \bibinfo {author} {\bibfnamefont {J.~D.}\ \bibnamefont
  {Swalen}},\ }\href {\doibase 10.1103/RevModPhys.31.841} {\bibfield  {journal}
  {\bibinfo  {journal} {Rev. Mod. Phys.}\ }\textbf {\bibinfo {volume} {31}},\
  \bibinfo {pages} {841} (\bibinfo {year} {1959})}\BibitemShut {NoStop}%
\bibitem [{\citenamefont {Herbst}\ and\ \citenamefont {van
  Dishoeck}(2009)}]{Herbst2009}%
  \BibitemOpen
  \bibfield  {author} {\bibinfo {author} {\bibfnamefont {E.}~\bibnamefont
  {Herbst}}\ and\ \bibinfo {author} {\bibfnamefont {E.~F.}\ \bibnamefont {van
  Dishoeck}},\ }\href {\doibase 10.1146/annurev-astro-082708-101654} {\bibfield
   {journal} {\bibinfo  {journal} {Ann. Rev. of Astron. Astroph.}\ }\textbf
  {\bibinfo {volume} {47}},\ \bibinfo {pages} {427} (\bibinfo {year}
  {2009})}\BibitemShut {NoStop}%
\bibitem [{\citenamefont {Muller}\ \emph {et~al.}(2011)\citenamefont {Muller},
  \citenamefont {Beelen}, \citenamefont {Gu\'elin}, \citenamefont {Aalto},
  \citenamefont {Black}, \citenamefont {Combes}, \citenamefont {Curran},
  \citenamefont {Theule},\ and\ \citenamefont {Longmore}}]{muller2011}%
  \BibitemOpen
  \bibfield  {author} {\bibinfo {author} {\bibfnamefont {S.}~\bibnamefont
  {Muller}}, \bibinfo {author} {\bibfnamefont {A.}~\bibnamefont {Beelen}},
  \bibinfo {author} {\bibfnamefont {M.}~\bibnamefont {Gu\'elin}}, \bibinfo
  {author} {\bibfnamefont {S.}~\bibnamefont {Aalto}}, \bibinfo {author}
  {\bibfnamefont {J.~H.}\ \bibnamefont {Black}}, \bibinfo {author}
  {\bibfnamefont {F.}~\bibnamefont {Combes}}, \bibinfo {author} {\bibfnamefont
  {S.}~\bibnamefont {Curran}}, \bibinfo {author} {\bibfnamefont
  {P.}~\bibnamefont {Theule}}, \ and\ \bibinfo {author} {\bibfnamefont
  {S.}~\bibnamefont {Longmore}},\ }\href {http://arxiv.org/abs/1104.3361v1}
  {\bibfield  {journal} {\bibinfo  {journal} {arXiv:1104.3361v1 [astro-ph.CO]}\
  } (\bibinfo {year} {2011})}\BibitemShut {NoStop}%
\bibitem [{\citenamefont {Henkel}()}]{Henkel2011}%
  \BibitemOpen
  \bibfield  {author} {\bibinfo {author} {\bibfnamefont {C.}~\bibnamefont
  {Henkel}},\ }\href@noop {} {}\bibinfo {howpublished} {Private
  Communication}\BibitemShut {NoStop}%
\bibitem [{\citenamefont {Ilyushin~\emph{et al.}}()}]{Ilyushin2011}%
  \BibitemOpen
  \bibfield  {author} {\bibinfo {author} {\bibfnamefont {V.~V.}\ \bibnamefont
  {Ilyushin~\emph{et al.}}},\ }\href@noop {} {\bibinfo  {journal} {in
  preparation}\ }\BibitemShut {NoStop}%
\bibitem [{\citenamefont {Hougen}\ \emph {et~al.}(1994)\citenamefont {Hougen},
  \citenamefont {Kleiner},\ and\ \citenamefont {Godefroid}}]{Hougen1994}%
  \BibitemOpen
\bibfield  {journal} {  }\bibfield  {author} {\bibinfo {author} {\bibfnamefont
  {J.~T.}\ \bibnamefont {Hougen}}, \bibinfo {author} {\bibfnamefont
  {I.}~\bibnamefont {Kleiner}}, \ and\ \bibinfo {author} {\bibfnamefont
  {M.}~\bibnamefont {Godefroid}},\ }\href {\doibase DOI:
  10.1006/jmsp.1994.1047} {\bibfield  {journal} {\bibinfo  {journal} {J. Mol.
  Spectrosc.}\ }\textbf {\bibinfo {volume} {163}},\ \bibinfo {pages} {559 }
  (\bibinfo {year} {1994})},\ \bibinfo {note} {(program available through:
  \url{http://www.ifpan.edu.pl/~kisiel/introt/introt.htm#belgi})}\BibitemShut
  {NoStop}%
\bibitem [{\citenamefont {Kleiner}(2010)}]{Kleiner2010}%
  \BibitemOpen
  \bibfield  {author} {\bibinfo {author} {\bibfnamefont {I.}~\bibnamefont
  {Kleiner}},\ }\href {\doibase DOI: 10.1016/j.jms.2009.12.011} {\bibfield
  {journal} {\bibinfo  {journal} {J. Mol. Spectrosc.}\ }\textbf {\bibinfo
  {volume} {260}},\ \bibinfo {pages} {1 } (\bibinfo {year} {2010})}\BibitemShut
  {NoStop}%
\bibitem [{\citenamefont {Lees}\ and\ \citenamefont
  {Baker}(1968)}]{Lees&Baker1968}%
  \BibitemOpen
  \bibfield  {author} {\bibinfo {author} {\bibfnamefont {R.~M.}\ \bibnamefont
  {Lees}}\ and\ \bibinfo {author} {\bibfnamefont {J.~G.}\ \bibnamefont
  {Baker}},\ }\href {\doibase 10.1063/1.1668221} {\bibfield  {journal}
  {\bibinfo  {journal} {J. Chem. Phys.}\ }\textbf {\bibinfo {volume} {48}},\
  \bibinfo {pages} {5299} (\bibinfo {year} {1968})}\BibitemShut {NoStop}%
\bibitem [{\citenamefont {Herbst}\ \emph {et~al.}(1984)\citenamefont {Herbst},
  \citenamefont {Messer}, \citenamefont {Lucia},\ and\ \citenamefont
  {Helminger}}]{Herbst1984}%
  \BibitemOpen
  \bibfield  {author} {\bibinfo {author} {\bibfnamefont {E.}~\bibnamefont
  {Herbst}}, \bibinfo {author} {\bibfnamefont {J.~K.}\ \bibnamefont {Messer}},
  \bibinfo {author} {\bibfnamefont {F.~C.~D.}\ \bibnamefont {Lucia}}, \ and\
  \bibinfo {author} {\bibfnamefont {P.}~\bibnamefont {Helminger}},\ }\href
  {\doibase DOI: 10.1016/0022-2852(84)90285-6} {\bibfield  {journal} {\bibinfo
  {journal} {J. Mol. Spectrosc.}\ }\textbf {\bibinfo {volume} {108}},\ \bibinfo
  {pages} {42 } (\bibinfo {year} {1984})}\BibitemShut {NoStop}%
\bibitem [{\citenamefont {Ilyushin}\ \emph {et~al.}(2001)\citenamefont
  {Ilyushin}, \citenamefont {Alekseev}, \citenamefont {Dyubko}, \citenamefont
  {Podnos}, \citenamefont {Kleiner}, \citenamefont {Margulès}, \citenamefont
  {Wlodarczak}, \citenamefont {Demaison}, \citenamefont {Cosléou},
  \citenamefont {Maté}, \citenamefont {Karyakin}, \citenamefont
  {Golubiatnikov}, \citenamefont {Fraser}, \citenamefont {Suenram},\ and\
  \citenamefont {Hougen}}]{Ilyushin2001}%
  \BibitemOpen
  \bibfield  {author} {\bibinfo {author} {\bibfnamefont {V.~V.}\ \bibnamefont
  {Ilyushin}}, \bibinfo {author} {\bibfnamefont {E.~A.}\ \bibnamefont
  {Alekseev}}, \bibinfo {author} {\bibfnamefont {S.~F.}\ \bibnamefont
  {Dyubko}}, \bibinfo {author} {\bibfnamefont {S.~V.}\ \bibnamefont {Podnos}},
  \bibinfo {author} {\bibfnamefont {I.}~\bibnamefont {Kleiner}}, \bibinfo
  {author} {\bibfnamefont {L.}~\bibnamefont {Margulès}}, \bibinfo {author}
  {\bibfnamefont {G.}~\bibnamefont {Wlodarczak}}, \bibinfo {author}
  {\bibfnamefont {J.}~\bibnamefont {Demaison}}, \bibinfo {author}
  {\bibfnamefont {J.}~\bibnamefont {Cosléou}}, \bibinfo {author} {\bibfnamefont
  {B.}~\bibnamefont {Maté}}, \bibinfo {author} {\bibfnamefont {E.~N.}\
  \bibnamefont {Karyakin}}, \bibinfo {author} {\bibfnamefont {G.~Y.}\
  \bibnamefont {Golubiatnikov}}, \bibinfo {author} {\bibfnamefont {G.~T.}\
  \bibnamefont {Fraser}}, \bibinfo {author} {\bibfnamefont {R.~D.}\
  \bibnamefont {Suenram}}, \ and\ \bibinfo {author} {\bibfnamefont {J.~T.}\
  \bibnamefont {Hougen}},\ }\href {\doibase DOI: 10.1006/jmsp.2000.8270}
  {\bibfield  {journal} {\bibinfo  {journal} {J. Mol. Spectrosc.}\ }\textbf
  {\bibinfo {volume} {205}},\ \bibinfo {pages} {286 } (\bibinfo {year}
  {2001})}\BibitemShut {NoStop}%
\bibitem [{\citenamefont {Kirtman}(1962)}]{kirtman1962}%
  \BibitemOpen
  \bibfield  {author} {\bibinfo {author} {\bibfnamefont {B.}~\bibnamefont
  {Kirtman}},\ }\href {\doibase 10.1063/1.1733049} {\bibfield  {journal}
  {\bibinfo  {journal} {J. Chem. Phys.}\ }\textbf {\bibinfo {volume} {37}},\
  \bibinfo {pages} {2516} (\bibinfo {year} {1962})}\BibitemShut {NoStop}%
\bibitem [{\citenamefont {Lees}(1973)}]{Lees1973}%
  \BibitemOpen
  \bibfield  {author} {\bibinfo {author} {\bibfnamefont {R.~M.}\ \bibnamefont
  {Lees}},\ }\href@noop {} {\bibfield  {journal} {\bibinfo  {journal}
  {Astrophys. J.}\ }\textbf {\bibinfo {volume} {184}},\ \bibinfo {pages} {763}
  (\bibinfo {year} {1973})}\BibitemShut {NoStop}%
\bibitem [{\citenamefont {Xu}\ \emph {et~al.}(2008)\citenamefont {Xu},
  \citenamefont {Fisher}, \citenamefont {Lees}, \citenamefont {Shi},
  \citenamefont {Hougen}, \citenamefont {Pearson}, \citenamefont {Drouin},
  \citenamefont {Blake},\ and\ \citenamefont {Braakman}}]{Xu2008o}%
  \BibitemOpen
  \bibfield  {author} {\bibinfo {author} {\bibfnamefont {L.-H.}\ \bibnamefont
  {Xu}}, \bibinfo {author} {\bibfnamefont {J.}~\bibnamefont {Fisher}}, \bibinfo
  {author} {\bibfnamefont {R.}~\bibnamefont {Lees}}, \bibinfo {author}
  {\bibfnamefont {H.}~\bibnamefont {Shi}}, \bibinfo {author} {\bibfnamefont
  {J.}~\bibnamefont {Hougen}}, \bibinfo {author} {\bibfnamefont
  {J.}~\bibnamefont {Pearson}}, \bibinfo {author} {\bibfnamefont
  {B.}~\bibnamefont {Drouin}}, \bibinfo {author} {\bibfnamefont
  {G.}~\bibnamefont {Blake}}, \ and\ \bibinfo {author} {\bibfnamefont
  {R.}~\bibnamefont {Braakman}},\ }\href {\doibase DOI:
  10.1016/j.jms.2008.03.017} {\bibfield  {journal} {\bibinfo  {journal} {J.
  Mol. Spectrosc.}\ }\textbf {\bibinfo {volume} {251}},\ \bibinfo {pages} {305
  } (\bibinfo {year} {2008})}\BibitemShut {NoStop}%
\bibitem [{\citenamefont {Kleiner}\ \emph {et~al.}(1996)\citenamefont
  {Kleiner}, \citenamefont {Hougen}, \citenamefont {Grabow}, \citenamefont
  {Belov}, \citenamefont {Tretyakov},\ and\ \citenamefont
  {Cosléou}}]{Kleiner1996}%
  \BibitemOpen
  \bibfield  {author} {\bibinfo {author} {\bibfnamefont {I.}~\bibnamefont
  {Kleiner}}, \bibinfo {author} {\bibfnamefont {J.~T.}\ \bibnamefont {Hougen}},
  \bibinfo {author} {\bibfnamefont {J.~U.}\ \bibnamefont {Grabow}}, \bibinfo
  {author} {\bibfnamefont {S.~P.}\ \bibnamefont {Belov}}, \bibinfo {author}
  {\bibfnamefont {M.~Y.}\ \bibnamefont {Tretyakov}}, \ and\ \bibinfo {author}
  {\bibfnamefont {J.}~\bibnamefont {Cosléou}},\ }\href {\doibase DOI:
  10.1006/jmsp.1996.0182} {\bibfield  {journal} {\bibinfo  {journal} {J. Mol.
  Spectrosc.}\ }\textbf {\bibinfo {volume} {179}},\ \bibinfo {pages} {41 }
  (\bibinfo {year} {1996})}\BibitemShut {NoStop}%
\bibitem [{\citenamefont {Ilyushin}\ \emph {et~al.}(2008)\citenamefont
  {Ilyushin}, \citenamefont {Kleiner},\ and\ \citenamefont
  {Lovas}}]{Ilyushin2008}%
  \BibitemOpen
  \bibfield  {author} {\bibinfo {author} {\bibfnamefont {V.}~\bibnamefont
  {Ilyushin}}, \bibinfo {author} {\bibfnamefont {I.}~\bibnamefont {Kleiner}}, \
  and\ \bibinfo {author} {\bibfnamefont {F.~J.}\ \bibnamefont {Lovas}},\ }\href
  {\doibase 10.1063/1.2815328} {\bibfield  {journal} {\bibinfo  {journal} {J.
  Phys. Chem. Ref. Data}\ }\textbf {\bibinfo {volume} {37}},\ \bibinfo {pages}
  {97} (\bibinfo {year} {2008})}\BibitemShut {NoStop}%
\bibitem [{\citenamefont {Ubachs}\ \emph {et~al.}(2007)\citenamefont {Ubachs},
  \citenamefont {Buning}, \citenamefont {Eikema},\ and\ \citenamefont
  {Reinhold}}]{Ubachs2007}%
  \BibitemOpen
  \bibfield  {author} {\bibinfo {author} {\bibfnamefont {W.}~\bibnamefont
  {Ubachs}}, \bibinfo {author} {\bibfnamefont {R.}~\bibnamefont {Buning}},
  \bibinfo {author} {\bibfnamefont {K.}~\bibnamefont {Eikema}}, \ and\ \bibinfo
  {author} {\bibfnamefont {E.}~\bibnamefont {Reinhold}},\ }\href {\doibase DOI:
  10.1016/j.jms.2006.12.004} {\bibfield  {journal} {\bibinfo  {journal} {J.
  Mol. Spectrosc.}\ }\textbf {\bibinfo {volume} {241}},\ \bibinfo {pages} {155
  } (\bibinfo {year} {2007})}\BibitemShut {NoStop}%
\bibitem [{\citenamefont {Lovas}(2004)}]{Lovas2004}%
  \BibitemOpen
  \bibfield  {author} {\bibinfo {author} {\bibfnamefont {F.~J.}\ \bibnamefont
  {Lovas}},\ }\href {\doibase 10.1063/1.1633275} {\bibfield  {journal}
  {\bibinfo  {journal} {J. Phys. Chem. Ref. Data}\ }\textbf {\bibinfo {volume}
  {33}},\ \bibinfo {pages} {177} (\bibinfo {year} {2004})}\BibitemShut
  {NoStop}%
\bibitem [{\citenamefont {Ilyushin}\ \emph {et~al.}(2004)\citenamefont
  {Ilyushin}, \citenamefont {Alekseev}, \citenamefont {Dyubko}, \citenamefont
  {Kleiner},\ and\ \citenamefont {Hougen}}]{Ilyushin2004}%
  \BibitemOpen
  \bibfield  {author} {\bibinfo {author} {\bibfnamefont {V.}~\bibnamefont
  {Ilyushin}}, \bibinfo {author} {\bibfnamefont {E.}~\bibnamefont {Alekseev}},
  \bibinfo {author} {\bibfnamefont {S.}~\bibnamefont {Dyubko}}, \bibinfo
  {author} {\bibfnamefont {I.}~\bibnamefont {Kleiner}}, \ and\ \bibinfo
  {author} {\bibfnamefont {J.}~\bibnamefont {Hougen}},\ }\href {\doibase DOI:
  10.1016/j.jms.2004.05.014} {\bibfield  {journal} {\bibinfo  {journal} {J.
  Mol. Spectrosc.}\ }\textbf {\bibinfo {volume} {227}},\ \bibinfo {pages} {115
  } (\bibinfo {year} {2004})}\BibitemShut {NoStop}%
\bibitem [{\citenamefont {Carvajal}\ \emph {et~al.}(2007)\citenamefont
  {Carvajal}, \citenamefont {Willaert}, \citenamefont {Demaison},\ and\
  \citenamefont {Kleiner}}]{Carvajal2007}%
  \BibitemOpen
  \bibfield  {author} {\bibinfo {author} {\bibfnamefont {M.}~\bibnamefont
  {Carvajal}}, \bibinfo {author} {\bibfnamefont {F.}~\bibnamefont {Willaert}},
  \bibinfo {author} {\bibfnamefont {J.}~\bibnamefont {Demaison}}, \ and\
  \bibinfo {author} {\bibfnamefont {I.}~\bibnamefont {Kleiner}},\ }\href
  {\doibase DOI: 10.1016/j.jms.2007.08.009} {\bibfield  {journal} {\bibinfo
  {journal} {J. Mol. Spectrosc.}\ }\textbf {\bibinfo {volume} {246}},\ \bibinfo
  {pages} {158 } (\bibinfo {year} {2007})}\BibitemShut {NoStop}%
\bibitem [{\citenamefont {Levshakov}\ \emph {et~al.}(2011)\citenamefont
  {Levshakov}, \citenamefont {Kozlov},\ and\ \citenamefont
  {Reimers}}]{Levshakov2011}%
  \BibitemOpen
  \bibfield  {author} {\bibinfo {author} {\bibfnamefont {S.~A.}\ \bibnamefont
  {Levshakov}}, \bibinfo {author} {\bibfnamefont {M.~G.}\ \bibnamefont
  {Kozlov}}, \ and\ \bibinfo {author} {\bibfnamefont {D.}~\bibnamefont
  {Reimers}},\ }\href {http://stacks.iop.org/0004-637X/738/i=1/a=26} {\bibfield
   {journal} {\bibinfo  {journal} {Astrophys. J.}\ }\textbf {\bibinfo {volume}
  {738}},\ \bibinfo {pages} {26} (\bibinfo {year} {2011})}\BibitemShut
  {NoStop}%
\bibitem [{\citenamefont {Kozlov}(2011)}]{Kozlov2011b}%
  \BibitemOpen
  \bibfield  {author} {\bibinfo {author} {\bibfnamefont {M.~G.}\ \bibnamefont
  {Kozlov}},\ }\href {http://arxiv.org/abs/1108.4520v1} {\bibfield  {journal}
  {\bibinfo  {journal} {arXiv:1108.4520v1 [astro-ph.CO]}\ } (\bibinfo {year}
  {2011})}\BibitemShut {NoStop}%
\bibitem [{\citenamefont {Malec}\ \emph {et~al.}(2010)\citenamefont {Malec},
  \citenamefont {Buning}, \citenamefont {Murphy}, \citenamefont {Milutinovic},
  \citenamefont {Ellison}, \citenamefont {Prochaska}, \citenamefont {Kaper},
  \citenamefont {Tumlinson}, \citenamefont {Carswell},\ and\ \citenamefont
  {Ubachs}}]{Malec2010}%
  \BibitemOpen
  \bibfield  {author} {\bibinfo {author} {\bibfnamefont {A.~L.}\ \bibnamefont
  {Malec}}, \bibinfo {author} {\bibfnamefont {R.}~\bibnamefont {Buning}},
  \bibinfo {author} {\bibfnamefont {M.~T.}\ \bibnamefont {Murphy}}, \bibinfo
  {author} {\bibfnamefont {N.}~\bibnamefont {Milutinovic}}, \bibinfo {author}
  {\bibfnamefont {S.~L.}\ \bibnamefont {Ellison}}, \bibinfo {author}
  {\bibfnamefont {J.~X.}\ \bibnamefont {Prochaska}}, \bibinfo {author}
  {\bibfnamefont {L.}~\bibnamefont {Kaper}}, \bibinfo {author} {\bibfnamefont
  {J.}~\bibnamefont {Tumlinson}}, \bibinfo {author} {\bibfnamefont {R.~F.}\
  \bibnamefont {Carswell}}, \ and\ \bibinfo {author} {\bibfnamefont
  {W.}~\bibnamefont {Ubachs}},\ }\href {\doibase
  10.1111/j.1365-2966.2009.16227.x} {\bibfield  {journal} {\bibinfo  {journal}
  {Monthly Notices of the Royal Astronomical Society}\ }\textbf {\bibinfo
  {volume} {403}},\ \bibinfo {pages} {1541} (\bibinfo {year}
  {2010})}\BibitemShut {NoStop}%
\bibitem [{\citenamefont {van Weerdenburg}\ \emph {et~al.}(2011)\citenamefont
  {van Weerdenburg}, \citenamefont {Murphy}, \citenamefont {Malec},
  \citenamefont {Kaper},\ and\ \citenamefont {Ubachs}}]{vanWeerdenburg2011}%
  \BibitemOpen
  \bibfield  {author} {\bibinfo {author} {\bibfnamefont {F.}~\bibnamefont {van
  Weerdenburg}}, \bibinfo {author} {\bibfnamefont {M.~T.}\ \bibnamefont
  {Murphy}}, \bibinfo {author} {\bibfnamefont {A.~L.}\ \bibnamefont {Malec}},
  \bibinfo {author} {\bibfnamefont {L.}~\bibnamefont {Kaper}}, \ and\ \bibinfo
  {author} {\bibfnamefont {W.}~\bibnamefont {Ubachs}},\ }\href {\doibase
  10.1103/PhysRevLett.106.180802} {\bibfield  {journal} {\bibinfo  {journal}
  {Phys. Rev. Lett.}\ }\textbf {\bibinfo {volume} {106}},\ \bibinfo {pages}
  {180802} (\bibinfo {year} {2011})}\BibitemShut {NoStop}%
\bibitem [{\citenamefont {Murphy}\ \emph {et~al.}(2008)\citenamefont {Murphy},
  \citenamefont {Flambaum}, \citenamefont {Muller},\ and\ \citenamefont
  {Henkel}}]{Murphy2008}%
  \BibitemOpen
  \bibfield  {author} {\bibinfo {author} {\bibfnamefont {M.~T.}\ \bibnamefont
  {Murphy}}, \bibinfo {author} {\bibfnamefont {V.~V.}\ \bibnamefont
  {Flambaum}}, \bibinfo {author} {\bibfnamefont {S.}~\bibnamefont {Muller}}, \
  and\ \bibinfo {author} {\bibfnamefont {C.}~\bibnamefont {Henkel}},\ }\href
  {\doibase 10.1126/science.1156352} {\bibfield  {journal} {\bibinfo  {journal}
  {Science}\ }\textbf {\bibinfo {volume} {320}},\ \bibinfo {pages} {1611}
  (\bibinfo {year} {2008})}\BibitemShut {NoStop}%
\bibitem [{\citenamefont {Henkel}\ \emph {et~al.}(2009)\citenamefont {Henkel},
  \citenamefont {Menten}, \citenamefont {Murphy}, \citenamefont {Jethava},
  \citenamefont {Flambaum}, \citenamefont {Braatz}, \citenamefont {Muller},
  \citenamefont {Ott},\ and\ \citenamefont {Mao}}]{Henkel2009}%
  \BibitemOpen
  \bibfield  {author} {\bibinfo {author} {\bibfnamefont {C.}~\bibnamefont
  {Henkel}}, \bibinfo {author} {\bibfnamefont {K.~M.}\ \bibnamefont {Menten}},
  \bibinfo {author} {\bibfnamefont {M.~T.}\ \bibnamefont {Murphy}}, \bibinfo
  {author} {\bibfnamefont {N.}~\bibnamefont {Jethava}}, \bibinfo {author}
  {\bibfnamefont {V.~V.}\ \bibnamefont {Flambaum}}, \bibinfo {author}
  {\bibfnamefont {J.~A.}\ \bibnamefont {Braatz}}, \bibinfo {author}
  {\bibfnamefont {S.}~\bibnamefont {Muller}}, \bibinfo {author} {\bibfnamefont
  {J.}~\bibnamefont {Ott}}, \ and\ \bibinfo {author} {\bibfnamefont {R.~Q.}\
  \bibnamefont {Mao}},\ }\href {\doibase 10.1051/0004-6361/200811475}
  {\bibfield  {journal} {\bibinfo  {journal} {Astron. Astrophys}\ }\textbf
  {\bibinfo {volume} {500}},\ \bibinfo {pages} {725} (\bibinfo {year}
  {2009})}\BibitemShut {NoStop}%
\bibitem [{\citenamefont {Kanekar}(2011)}]{Kanekar2011}%
  \BibitemOpen
  \bibfield  {author} {\bibinfo {author} {\bibfnamefont {N.}~\bibnamefont
  {Kanekar}},\ }\href {http://stacks.iop.org/2041-8205/728/i=1/a=L12}
  {\bibfield  {journal} {\bibinfo  {journal} {Astroph. J. Lett.}\ }\textbf
  {\bibinfo {volume} {728}},\ \bibinfo {pages} {L12} (\bibinfo {year}
  {2011})}\BibitemShut {NoStop}%
\bibitem [{\citenamefont {Levshakov~\emph{et al.}}(2010)}]{Levshakov2010}%
  \BibitemOpen
  \bibfield  {author} {\bibinfo {author} {\bibfnamefont {S.~A.}\ \bibnamefont
  {Levshakov~\emph{et al.}}},\ }\href {\doibase 10.1051/0004-6361/200913007}
  {\bibfield  {journal} {\bibinfo  {journal} {Astron. Astrophys}\ }\textbf
  {\bibinfo {volume} {512}},\ \bibinfo {pages} {A44} (\bibinfo {year}
  {2010})}\BibitemShut {NoStop}%
\bibitem [{\citenamefont {van~de Meerakker}\ \emph {et~al.}(2008)\citenamefont
  {van~de Meerakker}, \citenamefont {Bethlem},\ and\ \citenamefont
  {Meijer}}]{vandeMeerakker2008}%
  \BibitemOpen
  \bibfield  {author} {\bibinfo {author} {\bibfnamefont {S.~Y.~T.}\
  \bibnamefont {van~de Meerakker}}, \bibinfo {author} {\bibfnamefont {H.~L.}\
  \bibnamefont {Bethlem}}, \ and\ \bibinfo {author} {\bibfnamefont
  {G.}~\bibnamefont {Meijer}},\ }\href {\doibase 10.1038/nphys1031} {\bibfield
  {journal} {\bibinfo  {journal} {Nat. Phys.}\ }\textbf {\bibinfo {volume}
  {4}},\ \bibinfo {pages} {595} (\bibinfo {year} {2008})}\BibitemShut {NoStop}%
\bibitem [{\citenamefont {Bethlem}\ \emph {et~al.}(2008)\citenamefont
  {Bethlem}, \citenamefont {Kajita}, \citenamefont {Sartakov}, \citenamefont
  {Meijer},\ and\ \citenamefont {Ubachs}}]{Bethlem2008}%
  \BibitemOpen
  \bibfield  {author} {\bibinfo {author} {\bibfnamefont {H.~L.}\ \bibnamefont
  {Bethlem}}, \bibinfo {author} {\bibfnamefont {M.}~\bibnamefont {Kajita}},
  \bibinfo {author} {\bibfnamefont {B.}~\bibnamefont {Sartakov}}, \bibinfo
  {author} {\bibfnamefont {G.}~\bibnamefont {Meijer}}, \ and\ \bibinfo {author}
  {\bibfnamefont {W.}~\bibnamefont {Ubachs}},\ }\href {\doibase
  10.1140/epjst/e2008-00809-5} {\bibfield  {journal} {\bibinfo  {journal} {Eur.
  Phys. J. Special Topics}\ }\textbf {\bibinfo {volume} {163}},\ \bibinfo
  {pages} {55} (\bibinfo {year} {2008})}\BibitemShut {NoStop}%
\end{thebibliography}
\end{document}